\documentclass[aps]{revtex4}
\usepackage{rotate}
\usepackage{epsfig}
\usepackage{psfrag}

\newcommand \be  {\begin{equation}}
\newcommand \bea {\begin{eqnarray} \nonumber }
\newcommand \ee  {\end{equation}}
\newcommand \eea {\end{eqnarray}}
\newcommand{\subs}[1]{{\mbox{\scriptsize #1}}}
\newcommand \Tr {\mbox{Tr}}
\newcommand \rmi {\mbox{i}}

\begin{document}

\title{Financial Applications of Random Matrix Theory: a short review}

\author{Jean-Philippe Bouchaud, Marc Potters}
\affiliation{Science \& Finance, Capital Fund Management, 6 Bd Haussmann, 75009 Paris France}

\begin{abstract}

\end{abstract}

\maketitle

\section{Introduction}

\subsection{Setting the stage}

\label{sect1.1}

The Mar\v{c}enko-Pastur 1967 paper \cite{MP} on the spectrum of empirical correlation matrices is both remarkable and
precocious. It turned out to be useful in many, very different contexts (neural networks, image processing, wireless 
communications, etc.) and was unknowingly rediscovered several times. Its primary aim, as a new statistical tool
to analyse large dimensional data sets, only became relevant in the last two decades, when the storage 
and handling of humongous data sets became routine in almost all fields -- physics, image analysis, genomics, epidemiology, 
engineering, economics and finance, to quote only a few. It is indeed very natural to try to identify common causes 
(or factors) that explain the dynamics of $N$ quantities. These quantities might be daily returns of the different 
stocks of the S\&P 500, monthly inflation
of different sectors of activity, motion of individual grains in a packed granular medium, or different biological 
indicators (blood pressure, cholesterol, etc.) within a population, etc., etc. (for reviews of other applications and techniques, see \cite{Verdu,Edelmann,IMJ1})
We will denote by $T$ the total number of observations of each of the $N$ quantities.
In the example of stock returns, $T$ is the total number of trading days in the sampled data; but in the biological
example, $T$ is the size of the population. The realization of the $i$th quantity ($i=1, \dots, N$) at ``time'' $t$ 
($t=1,\dots,T$) will be denoted $r_i^t$, which will be assumed in the following to be demeaned and standardized. The normalized $T \times N$ matrix of returns will be
denoted as ${\bf X}$: $X_{ti}=r_i^t/\sqrt{T}$. The simplest way to characterize the correlations between these quantities is to compute the Pearson estimator of
the correlation matrix:
\be
E_{ij} = \frac{1}{T} \sum_{t=1}^T \, r_i^t \, r_j^t \equiv \left({\bf X}^{{{T}}} {\bf X}\right)_{ij},
\ee
where ${\bf E}$ will denote the empirical correlation matrix (i.e. on a given realization), that one must carefully 
distinguish from the ``true'' correlation matrix ${\bf C}$ of the underlying statistical process (that might 
not even exist). In fact, the whole point of the Mar\v{c}enko-Pastur result is to characterize the difference between
${\bf E}$ and ${\bf C}$. Of course, if $N$ is small (say $N=4$) and the number of observations is large (say $T=10^6$), 
then we can intuitively expect that any observable computed using ${\bf E}$ will be very close to its ``true'' value,
computed using ${\bf C}$. For example, a consistent estimator of $\Tr {\bf C}^{-1}$ is given $\Tr {\bf E}^{-1}$ when
$T$ is large enough for a fixed $N$. This is the usual limit considered in statistics. However, in many applications
where $T$ is large, the number of observables $N$ is also large, such that the ratio $q=N/T$ is not very small compared to
one. We will find below that when $q$ is non zero, and for large $N$, $\Tr {\bf E}^{-1} =  \Tr {\bf C}^{-1}/(1-q)$. 
Typical number in the case of stocks is $N=500$ and $T=2500$, corresponding to 10 years of daily data, already quite a long 
strand compared to the lifetime of stocks or the expected structural evolution time of markets. For inflation indicators, 
20 years of monthly data produce a meager $T=240$, whereas the number of sectors of activity for which inflation is
recorded is around $N=30$. The relevant mathematical limit to focus on in these cases is $T \gg 1$, $N \gg 1$ but with
$q = N/T = O(1)$. The aim of this paper is to review several Random Matrix Theory (RMT) results that can be 
established in this special asymptotic limit, where the empirical density of eigenvalues (the spectrum) is strongly distorted 
when compared to the `true' density (corresponding to $q \to 0$). When $T \to \infty$, $N \to \infty$, the spectrum has
some degree of universality with respect to the distribution of the $r_i^t$'s; this makes RMT results particularly appealing. Although the
scope of these results is much broader (as alluded to above), we will gird our discussion to the applications
of RMT to financial markets, a topic about which a considerable number of papers have been devoted to in the last decade (see e.g. 
\cite{us1,us2,Krakow1,RSVD,BiroliStudent,Stanley1,Stanley2,Lillo1,Lillo2,Lillo3,Kondor1,Burda1,Burda2,Burda3,Malevergne,Guhr,Vivo,Marsili,Frahm,Zumbach,Bai})
The following mini-review is intended to guide the reader through various results that we consider to be important, with no
claim of being complete. We furthermore chose to state these results in a narrative style, rather than in a more rigorous Lemma-Theorem
fashion. We provide references where more precise statements can be found.

\subsection{Principal Component Analysis}
\label{sect1.2}
The correlation matrix defined above is by construction an $N \times N$ symmetric matrix, that can be diagonalized. This
is the basis of the well known Principal Component Analysis (PCA), aiming at decomposing the fluctuations of the quantity 
$r_i^t$ into decorrelated contributions (the `components') of decreasing variance. In terms of the eigenvalues 
$\lambda_\alpha$ and eigenvectors $\vec V_\alpha$, the decomposition reads:
\be\label{decomp}
r_i^t = \sum_{\alpha=1}^N \sqrt{\lambda_\alpha} V_{\alpha,i} \, \epsilon_\alpha^t
\ee
where $V_{\alpha,i}$ is the i-th component of $\vec V_\alpha$, and $\epsilon_\alpha^t$ are uncorrelated (for 
different $\alpha$'s) random variables of unit variance. Note that the $\epsilon_\alpha^t$ are not necessarily uncorrelated
in the ``time'' direction, and not necessarily Gaussian. This PCA decomposition is particularly useful when there 
is a strong separation between eigenvalues. For example if the largest eigenvalue $\lambda_1$ is much larger than all
the others, a good approximation of the dynamics of the $N$ variables $r_i$ reads:
\be
r_i^t \approx \sqrt{\lambda_1} V_{1,i} \, \epsilon_1^t,
\ee
in which case a single ``factor'' is enough to capture the phenomenon. When $N$ is fixed and $T \to \infty$, all the
eigenvalues and their corresponding eigenvectors can be trusted to extract meaningful information. As we will review in
detail below, this is not the case when $q =N/T = O(1)$, where only a subpart of the eigenvalue spectrum of the `true' matrix ${\bf C}$ 
can be reliably estimated. In fact, since ${\bf E}$ is by construction a sum of $T$ projectors, ${\bf E}$ has (generically) 
$(N - T)^+$ eigenvalues exactly equal to zero, corresponding to the $(N - T)^+$ dimensions not spanned by these $T$ projectors.
These zero eigenvalues are clearly spurious and do not correspond to anything real for ${\bf C}$.

It is useful to give early on a physical (or rather financial) interpretation of the eigenvectors $\vec V_\alpha$. The list of 
numbers $V_{\alpha,i}$ can be seen as the weights of the different stocks $i=1,\dots,N$ in a certain portfolio $\Pi_\alpha$,
where some stocks are `long' ($V_{\alpha,i} > 0$) while other are `short' ($V_{\alpha,i} < 0$). The realized risk ${\cal R}_\alpha^2$ of portfolio
$\Pi_\alpha$, as measured by the variance of its returns, is given by:
\be
{\cal R}_\alpha^2 = \frac1T \sum_t \left(\sum_i V_{\alpha,i} \, r_i^t\right)^2 = \sum_{ij} V_{\alpha,i} V_{\alpha,j} E_{ij} \equiv \lambda_\alpha.
\ee
The eigenvalue $\lambda_\alpha$ is therefore the risk of the investment in portfolio $\alpha$. Large eigenvalues correspond to a risky mix
of assets, whereas small eigenvalues correspond to a particularly quiet mix of assets. Typically, in stock markets, the largest eigenvalue
corresponds to investing roughly equally on all stocks: $V_{1,i}=1/\sqrt{N}$. This is called the `market mode' and is strongly correlated 
with the market index. There is no diversification in this portfolio: the only bet is whether the market as a whole will go up or down, this is
why the risk is large. Conversely, if two stocks move very tightly together 
(the canonical example would be Coca-cola and Pepsi), then buying one and selling the other leads to a portfolio that hardly moves, being only
sensitive to events that strongly differentiate the two companies. Correspondingly, there is a small eigenvalue of ${\bf E}$ with eigenvector close to
$(0,0,\dots,\sqrt{2}/2,0,\dots,\sqrt{2}/2,0,\dots,0,0)$, where the non zero components are localized on the pair of stocks.

A further property of the portfolios $\Pi_\alpha$ is that their returns are uncorrelated, since:
\be
\frac1T \sum_t \left(\sum_i V_{\alpha,i} \, r_i^t\right)\left(\sum_j V_{\beta,j} \, r_j^t\right)= \sum_{ij} V_{\alpha,i} V_{\beta,j} E_{ij} 
\equiv \lambda_\alpha \delta_{\alpha,\beta}.
\ee
The PCA of the correlation matrix therefore provides a list of `eigenportfolios', corresponding to uncorrelated investments with decreasing variance.

We should mention at this stage an interesting duality that, although trivial from a mathematical point of view, 
looks at first rather counter-intuitive. Instead of the $N \times N$ 
correlation matrix of the stock returns, one could define a $T \times T$ correlation matrix 
$\widetilde {\bf E}$ of the daily returns, as:
\be
{\widetilde E}^{tt'} = \frac1N \sum_i r_i^t r_i^{t'} = \frac{T}{N} {\bf X}{\bf X}^{{{T}}}.
\ee
This measures how similar day $t$ and day $t'$ are, in terms of the `pattern' created by the returns of the $N$ stocks. The duality we are speaking
about is that the non zero eigenvalues of $\bf{\widetilde E}$ and of ${\bf E}$ are precisely the same, up to a factor $T/N$. This is obvious from Eq. (\ref{decomp}),
where the $V_{\alpha,i}$ and the $\epsilon_\alpha^t$ play completely symmetric roles -- the fact that the $\epsilon_\alpha^t$ are uncorrelated for 
different $\alpha$'s means that these vectors of dimension $T$ are orthogonal, as are the $V_{\alpha,i}$. Using this decomposition, one indeed 
finds:
\be
{\widetilde E}^{tt'} =  \frac1N  \sum_\alpha  \lambda_\alpha \, \epsilon_\alpha^t \epsilon_\alpha^{t'},
\ee
showing that the non zero eigenvalues of $\bf{\widetilde E}$ are indeed $\lambda_\alpha$'s (up to a factor $1/q$). 
The corresponding eigenvectors of $\bf{\widetilde E}$ are simply the lists of the daily returns of the portfolios $\Pi_\alpha$. 
Of course, if $T > N$, $\bf{\widetilde E}$ has $T - N$ additional zero eigenvalues. 

\section{Return statistics and portfolio theory}

\subsection{Single asset returns: a short review}
\label{sect2.1}
Quite far from the simple assumption of textbook mathematical finance, the returns (i.e. the relative price changes) of any kind of traded financial instrument (stocks, 
currencies, interest rates, commodities, etc. \footnote{Even the implied volatility, which is the object traded in option markets and is the market
forecast for the future amplitude of the fluctuations, has daily returns that are described by the same type of anomalous statistics!}) are very far
from Gaussian. The unconditional distribution of returns has fat tails, decaying as a power law for large arguments. In fact, the empirical probability distribution function 
of returns on shortish time scales (say between a few minutes and a few days) can be reasonably well fitted by a Student-t distribution (see e.g. \cite{book}):\footnote{On longer time
scales, say weeks to months, the distribution approaches a Gaussian, albeit anomalously slowly (see \cite{book}).}
\be\label{Student}
P(r) = \frac{1}{\sqrt{\pi}} \frac{\Gamma(\frac{1+\mu}{2})}{\Gamma(\frac{\mu}{2})} \frac{a^\mu}{(r^2+a^2)^{\frac{1+\mu}{2}}}
\ee
where $a$ is a parameter related to the variance of the distribution through $\sigma^2=a^2/(\mu-2)$, and $\mu$ is in the range $3$ to $5$ \cite{Stanley}. We assume here and in the
following that the returns have zero mean, which is appropriate for short enough time scales: any long term drift is generally negligible compared to $\sigma$ for time scales up to a few weeks.

This unconditional distribution can however be misleading, since returns are in fact very far from IID random variables. In other words, the returns cannot be
thought of as independently drawn Student random variables. For one thing, such a model predicts that upon time aggregation, the distribution of returns 
is the convolution of Student distributions, which converges far too quickly towards a Gaussian distribution of returns for longer time scales. In
intuitive terms, the volatility of financial returns is itself a dynamical variable, that changes over time with a broad distribution of characteristic frequencies. 
In more formal terms, the return at time $t$ can be represented by the product of a volatility component $\sigma^t$ and a directional component $\xi^t$ (see e.g. \cite{book}):
\be\label{return}
r^t = \sigma^t \xi^t,
\ee
where the $\xi^t$ are IID random variables of unit variance, and $\sigma^t$ a positive random variable with both fast and slow components. It is to a large
extent a matter of
taste to choose $\xi^t$ to be Gaussian and keep a high frequency, unpredictable part to $\sigma^t$, or to choose $\xi^t$ to be non-Gaussian (for example 
Student-t distributed \footnote{A Student-t variable can indeed be written as $\sigma \xi$, where $\xi$ is Gaussian and $\sigma^2$ is an inverse Gamma random
variable, see below.}) and only keep the low frequency, predictable part of $\sigma^t$. The slow part of $\sigma^t$ is found to be a long memory process, such
that its correlation function decays as a slow power-law of the time lag $\tau$ (see \cite{book,MRW} and references therein):\footnote{The overline means an average over the volatility fluctuations, whereas
the brackets means an average over both the volatility ($\sigma^t$) and the directional ($\xi^t$) components.}
\be
\overline{\sigma^{t}\sigma^{t+\tau}} - \overline{\sigma}^2 \propto \tau^{-\nu}, \qquad \nu \sim 0.1
\ee
It is worth insisting that in Eq. (\ref{return}), $\sigma^t$ and $\xi^t$ are in fact not independent. It is indeed well documented that on stock markets 
negative past returns tend to increase future volatilities, and vice-versa \cite{book}. This is called the `leverage' effect, and means in particular that 
the average of quantities such as $\xi^t \sigma^{t+\tau}$ is negative when $\tau > 0$.

\subsection{Multivariate distribution of returns}
\label{sect2.2}
Having now specified the monovariate statistics of returns, we want to extend this description to the joint distribution of the returns of $N$ correlated 
assets. We will first focus on the joint distribution of {\it simultaneous} returns $\left\{r_1^t,r_2^t,\dots,r_N^t\right\}$. Clearly, all marginals of this
joint distribution must resemble the Student-t distribution (\ref{Student}) above; furthermore, it must be compatible with the (true) correlation matrix of the returns:
\be
C_{ij} = \int \prod_k \left[{\rm d}r_k\right] \, r_i r_j \, P(r_1,r_2,\dots,r_N).
\ee
Needless to say, these two requirements are weak constraints that can be fulfilled by the joint distribution $P(r_1,r_2,\dots,r_N)$ in an infinite number of ways. This is
referred to as the `copula specification problem' in quantitative finance. A copula is a joint distribution of $N$ random variables $u_i$ that all have a uniform marginal 
distribution in $[0,1]$; this can be transformed into $P(r_1,r_2,\dots,r_N)$ by transforming each $u_i$ into $r_i = F^{-1}_i(u_i)$, where $F_i$ is the (exact) cumulative marginal distribution
of $r_i$. The fact that the copula problem is hugely under-constrained has led to a proliferation of possible candidates for the structure of financial asset correlations (for a review,
see e.g. \cite{copula1,copula2,copula3,bookMS}). 
Unfortunately, the proposed copulas are often chosen because of mathematical convenience rather than based on a plausible underlying mechanism. From that point of view, many copulas 
appearing in the literature are in fact very unnatural. 

There is however a natural extension of the monovariate Student-t distribution that has a clear financial interpretation. If we generalize the above decomposition Eq. (\ref{return}) as:
\be\label{return2}
r_i^t = s_i \, \sigma^t \xi_i^t,
\ee
where the $\xi_i^t$ are correlated Gaussian random variables with a correlation matrix $\widehat C_{ij}$ and the volatility $\sigma^t$ is common to all assets and distributed as:
\be \label{defstudent}
P(\sigma)= \frac{2}{\Gamma(\frac{\mu}{2})}
\exp\left[-\frac{\sigma_0^2}{\sigma^2}\right] \frac{\sigma_0^\mu}{\sigma^{1+\mu}},
\ee
where $\sigma_0^2=2\mu/(\mu-2)$ in such a way that $\langle \sigma^2 \rangle=1$, such that $s_i$ is the volatility of the stock $i$. 
The joint distribution of returns is then a multivariate Student $P_S$ that reads explicitly:
\be \label{multivariateS2}
P_S(r_1, r_2,\dots, r_N)= \frac{\Gamma(\frac{N+\mu}{2})}{\Gamma(\frac{\mu}{2})
\sqrt{(\mu\pi)^N \det{\bf \widehat 
C}}} \frac{1}{\left(1+\frac{1}{\mu} \sum_{ij} r_i (\widehat C^{-1})_{ij} r_j \right)^{\frac{N+\mu}{2}}},
\ee
where we have normalized returns so that $s_i \equiv 1$. Let us list a few 
useful properties of this model:
\begin{itemize}
\item The marginal distribution of any $r_i$ is a monovariate Student-t 
distribution of parameter $\mu$.
\item In the limit $\mu \to \infty$, one can show that the multivariate 
Student distribution
$P_S$ tends towards a multivariate Gaussian distribution. This is expected, 
since in this limit, the random volatility $\sigma$ does not fluctuate anymore and is equal to $1$.
\item The correlation matrix of the $r_i$ is given, for $\mu > 2$, by:
\be
C_{ij} = \langle r_i r_j \rangle = \frac{\mu}{\mu-2} \widehat C_{ij}.
\ee
\item Wick's theorem for Gaussian variables can be extended to 
Student variables. For example, one can show that:
\be\label{Wick}
\langle r_i r_j r_k r_l \rangle = \frac{\mu-2}{\mu-4} \left[C_{ij}C_{kl}+C_{ik}C_{jl}+C_{il}C_{jk}\right],
\ee
This shows explicitly that uncorrelated by Student variables are not independent. Indeed, even when $C_{ij}=0$, the
correlation of squared returns is positive:
\be
\langle r_i^2 r_j^2 \rangle - \langle r_i^2 \rangle^2  =  \frac{2}{\mu-4} C_{ii}C_{jj} > 0.
\ee
\item Finally, note the matrix $\widehat C_{ij}$ can be estimated from empirical
data using a maximum likelihood procedure. Given a time series of stock returns
$r_i^t$, the most likely matrix $\widehat C_{ij}$ is given by the solution of the following equation:
\be
\label{maxlstudent}
\widehat C_{ij}=\frac{N+\mu}{T}\, \sum_{t=1}^T \frac{r_i^t r_j^t}
{\mu+\sum_{mn} r_m^t (\widehat C^{-1})_{mn} r_n^t}.
\ee
Note that in the Gaussian limit $\mu \to \infty$ for a fixed $N$, the denominator of the
above expression is simply given by $\mu$, and the final expression is 
simply:
\be
\widehat C_{ij}= C_{ij}=\frac{1}{T} \sum_{t=1}^T {r_i^t r_j^t},
\ee
as it should be.
\end{itemize}

This multivariate Student model is in fact too simple to describe financial data since it assumes that there is a unique volatility factor, common to all assets. One expects that in reality several
volatility factors are needed. However, the precise implementation of this idea and the resulting form of the multivariate distribution (and the corresponding natural copula) has not been worked out in details and
is still very much a research topic.

Before leaving this section, we should mention the role of the observation frequency, i.e. the time lag used to define  price returns. 
Qualitatively, all the above discussion applies as soon as one can forget about price discretization
effects (a few minutes on actively traded stocks) up to a few days, before a progressive `gaussianization' of returns takes place. Quantitatively, however, some measurable evolution with the
time lag can be observed. One important effect for our purpose here is the so-called Epps effect, i.e. the fact that the correlation of returns $r_i$ and $r_j$ tends to increase with the time lag, 
quite strongly between $5$ minutes and $30$ minutes, then more slowly before apparently saturating after a few days \cite{Epps, MantegnaTree}. A simple mechanism for such an increase (apart from artefacts coming from
microstructural effects and stale prices) is pair trading. Imagine two stocks $i$ and $j$ known to be similar to each other (say, as mentioned above, Coca and Pepsi). Then the evolution of 
one stock, due to some idiosyncratic effect, is expected to drive the other through the impact of pair traders. One can write down a mathematical model for this, and compute the lag dependence of 
the returns, but it is quite clear that the time scale over which the correlation coefficient converges towards its low frequency value is directly related to the strength of the pair trading effect. 

The Epps effect is very important since one might have hoped that increasing the frequency of observations allows one to have effectively longer samples of returns to estimate the correlation 
matrix, thereby increasing the quality factor $Q=T/N$ alluded to in the introduction. One has to make sure, however, that the very object one wants to measure, i.e. the matrix $C_{ij}$, does
not actually change with the observation frequency. It seems that the Epps effect is nowadays weaker than in the past 
(say before year 2000) in the sense that the correlation matrix converges faster towards its low frequency limit. But as we discuss in the conclusion, there might be a lot to learn from
a detailed analysis of the ultra high frequency behaviour of the correlation matrix.

\subsection{Risk and portfolio theory}
\label{sect2.3}
Suppose one builds a portfolio of $N$ assets with weight $w_i$ on the $i$th asset, with (daily) volatility $s_i$. 
If one knew the `true' correlation matrix $C_{ij}$, one would have access to the (daily) variance of the portfolio return,
given by:
\be\label{risk}
{\cal R}^2=\sum_{ij} w_i s_i C_{ij} s_j w_j,
\ee
where $C_{ij}$ is the correlation matrix.
If one has predicted gains $g_i$, then the expected gain of the portfolio is ${\cal G}=\sum w_i g_i$.

In order to measure and optimize the risk of this portfolio, one therefore has to come up with a 
reliable estimate of the correlation matrix $C_{ij}$. This is difficult in general since one has to 
determine of the order of $N^2/2$ coefficients out of $N$ time series of length $T$, and in general 
$T$ is not much larger than $N$. As noted in the introduction, typical values of $Q=T/N$ are in the range $1 \to 10$ in most applications.
In the following we assume for simplicity that the volatilities $s_i$ are perfectly known (an improved estimate of the future 
volatility over some time horizon can be obtained using the information distilled by option markets). By redefining $w_i$ as $w_i s_i$ 
and $g_i$ as $g_i/s_i$, one can set $s_i \equiv 1$, which is our convention from now on.

The risk of a portfolio with weights $w_i$ constructed {\it independently} of the past
realized returns $r_i^t$ is faithfully measured by:
\be
{\cal R}^2_E = \sum_{ij} w_i E_{ij} w_j,
\ee
using the empirical correlation matrix $\mathbf{E}$. This estimate is unbiased and the relative 
mean square-error one the risk is small ($\sim 1/T$). But when the $w$ are chosen using the observed $r$'s, as
we show now, the result can be very different.

Problems indeed arise when one wants to estimate the risk of an optimized portfolio, resulting from a 
Markowitz optimization scheme, which gives the portfolio with maximum expected return for a given risk or equivalently,
the minimum risk for a given return $\cal G$ (we will study the latter case below). Assuming ${\bf C}$ is known, simple calculations using Lagrange multipliers readily 
yield the optimal weights $w_i^*$, which read, in matrix notation:
\be\label{Msolution}
\mathbf{w}^*_C = {\cal G} \frac{\mathbf{C}^{-1}\mathbf{g}}{\mathbf{g}^{{{T}}}\mathbf{C}^{-1}\mathbf{g}}
\ee
One sees that these optimal weights involve the inverse of the correlation matrix, which will be the source 
of problems, and will require a way to `clean' the empirical correlation matrix. Let us explain why in details.

The question is to estimate the risk of this optimized portfolio, and in particular to understand the 
biases of different possible estimates. We define the following three quantities \cite{Krakow1}:
\begin{itemize} 
\item The ``in-sample'' risk, corresponding to the risk of the optimal portfolio over the period used
to construct it, using ${\bf E}$ as the correlation matrix.
\be
{\cal R}^2_\subs{in}=\mathbf{w}_E^{*{{T}}}\mathbf{E}\mathbf{w}^*_E = \frac{{\cal G}^2}{\mathbf{g}^{{{T}}}\mathbf{E}^{-1}\mathbf{g}}
\ee
\item The ``true'' minimal risk, which is the risk of the optimized portfolio in the ideal world where $\mathbf{C}$
is perfectly known:
\be
{\cal R}^2_\subs{true}=\mathbf{w}_C^{*{{T}}}\mathbf{C}\mathbf{w}^*_C =\frac{{\cal G}^2}{\mathbf{g}^{{{T}}}\mathbf{C}^{-1}\mathbf{g}}
\ee
\item  The ``out-of-sample'' risk which is the risk of the portfolio constructed using $\mathbf{E}$, 
but observed on the next (independent) period of time. The expected risk is then: 
\be
{\cal R}^2_\subs{out}=\mathbf{w}_E^{*{{T}}}\mathbf{C}\mathbf{w}^*_E=
\frac{{\cal G}^2\mathbf{g}^{{{T}}}\mathbf{E}^{-1}\mathbf{CE}^{-1}\mathbf{g}}{(\mathbf{g}^{{{T}}}\mathbf{E}^{-1}\mathbf{g})^2}
\ee
This last quantity is obviously the most important one in practice.
\end{itemize}
If we assume that $\mathbf{E}$ is a noisy, but unbiased estimator of $\mathbf{C}$, such that $\overline{\mathbf{E}}=\mathbf{C}$, 
one can use a convexity argument for the inverse of positive definite matrices to show that in general:
\be
\overline{\mathbf{g}^{{{T}}}\mathbf{E}^{-1}\mathbf{g}} \geq \mathbf{g}^{{{T}}}\mathbf{C}^{-1}\mathbf{g}
\ee
Hence for large matrices, for which the result is self-averaging:
\be
{\cal R}^2_\subs{in} \leq {\cal R}^2_\subs{true}.
\ee
By optimality, one clearly has:
\be
{\cal R}^2_\subs{true} \leq {\cal R}^2_\subs{out}.
\ee
These results show that the out-of-sample risk of an optimized portfolio is larger (and in practice, much larger,
see section \ref{sect5.2} below) than the in-sample risk, which itself is an underestimate of the true minimal risk. This is a 
general situation: using past returns to optimize a strategy always leads to over-optimistic results because 
the optimization adapts to the particular realization of the noise, and is unstable in time. 
Using the Random Matrix results of the next sections, one can show that for IID returns, with an {\it arbitrary}
``true'' correlation matrix ${\bf C}$, the risk of large portfolios obeys: \cite{Kondor1}
\be\label{risks}
{\cal R}_\subs{in}={{\cal R}_\subs{true}}\sqrt{1-q}={\cal R}_\subs{out}{(1-q)}.
\ee
where $q=N/T=1/Q$. The out-of-sample risk is therefore $1/\sqrt{1-q}$ times larger than the true risk, while the in sample risk is 
$\sqrt{1-q}$ smaller than the true risk. This is a typical data snooping effect. Only in the limit $q \to 0$ will these risks coincide, 
which is expected since in this case the measurement noise disappears, and ${\bf E}={\bf C}$. In the limit $q \to 1$, on the other hand, 
the in-sample risk becomes zero since it becomes possible to find eigenvectors (portfolios) with exactly zero eigenvalues, i.e., zero in sample 
risk. The underestimation of the risk turns out to be even stronger in the case of a multivariate Student model for
returns \cite{BiroliStudent}. In any case, the optimal determination of the correlation matrix based on empirical should be such that the ratio 
${\cal R}^2_\subs{true}/{\cal R}^2_\subs{out} \leq 1$ is as large as possible.

In order to get some general intuition on how the Markowitz optimal portfolio might not be optimal at all, let us rewrite the solution 
Eq. (\ref{Msolution}) above in terms of eigenvalues and eigenvectors:
\be
w_i^* \propto \sum_{\alpha j} \lambda_\alpha^{-1} V_{\alpha,i} V_{\alpha,j} g_j \equiv 
g_i + \sum_{\alpha j} \left(\lambda_\alpha^{-1}-1\right) V_{\alpha,i} V_{\alpha,j} g_j 
\ee
The first term corresponds to the naive solution: one should invest proportionally to the expected gain 
(in units where $s_i=1$). The correction term means that the weights of eigenvectors with $\lambda_\alpha > 1$ must be reduced, 
whereas the weights of eigenvectors with $\lambda_\alpha < 1$ should be enhanced. The optimal Markowitz solution 
may allocate a substantial weight to small eigenvalues, which may be entirely dominated by measurement noise and
hence unstable. There several ways to clean the correlation matrix such as to tame these spurious small risk portfolios, in particular 
based on Random Matrix Theory ideas. We will come back to this point in Sect. \ref{sect5.2}

\subsection{Non equal time correlations and more general rectangular correlation matrices}
\label{SVD_intro}
\label{sect2.4}
The equal time correlation matrix $C_{ij}$ is clearly important for risk purposes, and also the understand the structure of the market, or more generally the `Principle 
Components' driving the process under consideration. A natural extension, very useful for prediction purposes, is to study a lagged correlation matrix between past
and future returns. Let us define $C_{ij}(\tau)$ as:
\be\label{Ctau}
C_{ij}(\tau) = \langle r_i^t r_j^{t+\tau} \rangle
\ee
such that $C_{ij}(\tau=0)=C_{ij}$ is the standard correlation coefficient. Whereas $C_{ij}$ is clearly a symmetric matrix, $C_{ij}(\tau > 0)$ is in general non symmetric, 
and only obeys $C_{ij}(\tau)=C_{ji}(-\tau)$. How does one extend the idea of `Principle Components', seemingly associated to the diagonalisation of $C_{ij}$, to these 
assymetric case? 

The most general case looks in fact even worse: one could very well measure the correlation between $N$ `input' variables $X_i$, $i=1,...,N$
and $M$ `output' variables $Y_a$, $a=1,...,M$. The $X$ and the $Y$'s may be completely different from one another (for example, $X$ could be 
production indicators and $Y$ inflation indexes), or, as in the above example the same set of observables but observed at different times: $N=M$, 
$X_i^t = r_i^t$ and $Y_{a}^t=r_a^{t+\tau}$. The cross-correlations between $X$'s and $Y$'s is characterized by a rectangular $N \times M$ matrix
${\cal C}$ defined as:
\be
{\cal C}_{ia} = \langle X_i Y_a \rangle
\ee
(we assume that both $X$'s and $Y$'s have zero mean and variance unity). If there is a total of $T$ observations, where both $X_i^t$ and $Y_a^t$, $t=1,...,T$ are observed,
the empirical estimate of ${\cal C}$ is, after standardizing $X$ and $Y$:
\be\label{empiricalSVD}
{\cal E}_{ia} = \frac1T \sum_{t=1}^T X_i^t Y_a^t.
\ee

What can be said about these rectangular, non symmetric correlation matrices? The singular value decomposition (SVD) answers the question in the
following sense: what is the (normalized) linear combination of $X$'s on the one hand, and of $Y$'s on the other hand, that have the strongest 
mutual correlation? In other words, what is the best pair of predictor and predicted variables, given the data?
The largest singular value $c_{\max}$ and its corresponding left and right eigenvectors answer precisely this question: the
eigenvectors tell us how to construct these optimal linear combinations, and the associated singular value gives us
the strength of the cross-correlation: $0 \leq c_{\max} \leq 1$. One can now restrict both the input and output spaces to the $N-1$ and 
$M-1$ dimensional sub-spaces orthogonal to the two eigenvectors, and repeat the operation. The list of 
singular values $c_a$ gives the prediction power, in decreasing order, of the corresponding linear combinations. This is called ``Canonical Component Analysis'' (CCA)
in the literature \cite{IMJ2}; surprisingly in view of its wide range of applications, this method of investigation has be somewhat neglected since it was first introduced in 1936 \cite{Hotelling}. 

How to get these singular values and the associated left and right eigenvectors? The trick is to consider the $N \times N$ 
matrix ${\cal CC}^{{{T}}}$, which is now symmetric and has $N$ non negative eigenvalues, each of which being equal to the 
square of a singular value of ${\cal C}$ itself. The eigenvectors give us the weights of the linear combination of the $X$'s that
construct the `best' predictors in the above sense. One then forms the $M \times M$ matrix ${\cal C}^{{{T}}}{\cal C}$ that has 
exactly the same non zero eigenvalues as ${\cal CC}^{{{T}}}$; the corresponding eigenvectors now give us the weights of the linear combination of the 
$Y$'s that construct the `best' predictees. If $M > N$, ${\cal C}^{{{T}}}{\cal C}$ has $M-N$ additional zero eigenvalues; whereas when $M < N$ it is
${\cal CC}^{{{T}}}$ that has an excess of $N-M$ zero eigenvalues. The list of the non zero eigenvalues, $c_{\max}^2=c_1^2 \geq c_2^2 \geq \dots \geq c_{(N,M)^-}^2$ gives
a sense of the predictive power of the $X$'s on the behaviour of the $Y$'s. However, as for standard correlation matrices, the empirical determination 
of ${\cal C}$ is often strewn with measurement noise and RMT will help sorting out grain from chaff, i.e. what is true information (the ``grain'') in the spectrum of ${\cal C}$ and what is 
presumably the ``chaff''.

\section{Random Matrix Theory: The Bulk}

\subsection{Preliminaries}
\label{sect3.1}
Random Matrix Theory (RMT) attempts to make statements about the statistics of the eigenvalues $\lambda_\alpha$ of large random matrices, in particular the density of 
eigenvalues ${\rho}(\lambda)$, defined as:
\be
\rho_N(\lambda) = \frac1N \sum_{\alpha=1}^N \delta\left(\lambda - \lambda_\alpha\right),
\ee
where $\lambda_\alpha$ are the eigenvalues of the $N \times N$ symmetric matrix ${\bf H}$ that belongs to the statistical ensemble under scrutiny. It is customary to introduce the 
the resolvent $G_{{H}}(z)$ of ${\bf H}$ (also called the Stieltjes transform), where $z$ is a complex number:
\be
G_{{H}}(z)=\frac{1}{N}\Tr\left[(z\mathbf{I}-\mathbf{H})^{-1}\right],
\ee
from which one can extract the spectrum as:
\be\label{im_part}
\rho_N(\lambda) = \lim_{\epsilon \to 0} \frac{1}{\pi} \Im\left(G_{{H}}(\lambda-\rmi\epsilon)\right).
\ee

In the limit where $N$ tends to infinity, it often (but not always) happens that the density of eigenvalues $\rho_N$ tends almost surely to a unique well defined density $\rho_\infty(\lambda)$.
This means that the density $\rho_N$ becomes independent of the specific realization of the matrix ${\bf H}$, provided ${\bf H}$ is a `typical' sample within its ensemble. This property, called
`ergodicity' or `self-averaging', is extremely important for practical applications since the asymptotic result $\rho_\infty(\lambda)$ can be used describe the eigenvalue density of a 
{\it single} instance. This is clearly one of the key of the success of RMT.  

Several other `transforms', beyond the resolvent $G(z)$, turn out to be useful for our purposes. One is the so-called `Blue function' $B(z)$, which is the functional inverse of $G(z)$, i.e.:
$B[G(z)]=G[B(z)]=z$. The R-transform is simply related to the Blue function through \cite{Verdu}:
\be
R(z) = B(z)- z^{-1}.
\ee
It is a simple exercise to show that $R(z)$ obeys the following property:
\be\label{Rscaling}
R_{a{H}}(z) = a R_{{H}}(az)
\ee
where $a$ is an arbitrary real number. Furthermore, $R(z)$ can be expanded for large $z$ as $R(z)=\sum_{k=1}^\infty c_k z^{k-1}$, where the coefficients $c_k$ can be thought of as cumulants (see below).
For example, $c_1 = \int {\rm d}\lambda \lambda \rho(\lambda)$. When $c_1=0$, $c_2= \int {\rm d}\lambda \lambda^2 \rho(\lambda)$.

The last object that we will need is more cumbersome. It is called the S-transform and is defined as follows \cite{Verdu}:
\be
S(z)= - \frac{1+z}{z} \eta^{-1}(1+z) \quad {\mbox{where}} \quad \eta(y) \equiv - \frac{1}{y}G\left(\frac{1}{y}\right).
\ee

In the following, we will review several RMT results on the bulk density of states $\rho_\infty(\lambda)$ that can be obtained using an amazingly efficient concept: matrix freeness \cite{Voiculescu}. The
various fancy transforms introduced above will then appear more natural.

\subsection{Free Matrices}
\label{sect3.2}
Freeness is the generalization to matrices of the concept of independence for random variables. 
Loosely speaking, two matrices ${\bf A}$ and ${\bf B}$ are mutually free if their eigenbasis 
are related to one another by a random rotation, or said differently if the eigenvectors of ${\bf A}$ and ${\bf B}$ are almost surely orthogonal. 
A more precise and comprehensive definition can be found in, e.g. \cite{Verdu}, but our simplified definition, and the following examples, will be sufficient for our purposes. 

Let us give two important examples of mutually free matrices. The first one is nearly trivial. Take two fixed matrices ${\bf A}$ and ${\bf B}$, and choose a 
certain rotation matrix ${\bf O}$ within the orthogonal group $O(N)$, uniformly over the Haar measure. Then ${\bf A}$ and ${\bf O}^{{{T}}}{\bf B}{\bf O}$ are 
mutually free. The second is more interesting, and still not very esoteric. Take two matrices ${\bf H}_1$ and ${\bf H}_2$ chosen independently within the GOE ensemble, i.e. the 
ensemble symmetric matrices such that all entries are IID Gaussian variables. Since the measure of this ensemble of random matrices is invariant under orthogonal transformation, 
it means that the rotation matrix ${\bf O_1}$ diagonalizing ${\bf H}_1$ is a random rotation matrix over $O(N)$ (this is actually a convenient numerical method to generate random rotation matrices). 
The rotation ${\bf O}_1^{{{T}}}{\bf O}_2$  from the eigenbasis of ${\bf H}_1$ to that of ${\bf H}_2$ is therefore also random, and ${\bf H}_1$ and ${\bf H}_2$ are mutually free. 
More examples will be encountered below.

Now, matrix freeness allows one to compute the spectrum of the sum of matrices, knowing the spectrum of each of the matrices, provided they are mutually
free. More precisely, if $R_{A}(z)$ and $R_{B}(z)$ are the R-transforms of two free matrices ${\bf A}$ and ${\bf B}$, then:
\be 
R_{A+B}(z)=R_{A}(z)+R_{B}(z)
\ee
This result clearly generalizes the convolution rule for sum of two independent random variables, for which the logarithm of the characteristic function 
is additive. Once $R_{A+B}(z)$ is known, one can in principle invert the R-transform to reach the eigenvalue density of ${\bf A+B}$

There is an analogous result for the product of non negative random matrices. In this case, the S-transform is multiplicative:
\be 
S_{A+B}(z)=S_{A}(z)S_{B}(z)
\ee

In the rest of this section, we will show how these powerful rules  allows one to establish very easily several well known eigenvalue
densities for large matrices, as well as some newer results.

\subsection{Application: Wigner and Mar\v{c}enko \& Pastur}
\label{sect3.3}
Let us start with the celebrated Wigner semi-circle for Gaussian Orthogonal matrices. As stated above, two such matrices  ${\bf H}_1$ and ${\bf H}_2$
are mutually free. Furthermore, because of the stability of Gaussian variables under addition, $({\bf H}_1+{\bf H}_2)/\sqrt{2}$ is in the same
ensemble. One therefore has:
\be
R_{\sqrt{2}{H}}(z)=R_{{H}_1+{H}_2}(z)=R_{{H}_1}(z)+R_{{H}_2}(z)=2R_{{H}}(z)
\ee
Using the result Eq. (\ref{Rscaling}) above with $a=\sqrt{2}$, one finds that $R(z)$ must obey:
\be
2R_{{H}}(z)= \sqrt{2}R_{{H}}(\sqrt{2}z) \longrightarrow R_{{H}}(z)=z
\ee
where we have assumed the standard normalization $\Tr {\bf H}^2=1$. One can check easily that $R(z)=z$ is indeed the R-transform of Wigner semi-circle. 
There is another nice Central Limit Theorem-like way of establishing this result. 
Suppose ${\bf H}_i$, $i=1, \dots, {\cal N}$ are `small' traceless random matrices, such that each element has a variance 
equal to $\epsilon^2$ with $\epsilon \to 0$. Expand their resolvent $G_i(z)$ in $1/z$:
$$
G(z)=\frac1z + 0 + \epsilon^2 \frac{1}{z^3} + O(\epsilon^3/z^4) \to \frac1z \approx G - \epsilon^2 G^3.
$$
Hence,
$$
B(z) \approx \frac{1}{z-\epsilon^2 z^3} \to R(z)=B(z)-\frac1z \approx \epsilon^2 z + O(\epsilon^3 z^2)
$$
Now if these ${\cal N}$ matrices are mutually free, with $\epsilon = {\cal N}^{-1/2}$ and ${\cal N} \to \infty$, then the R-transform of the sum of such matrices is:
$$
R(z)={\cal N}\epsilon^2 z + O({\cal N}\epsilon^3 z^2) \to_{{\cal N} \to \infty} z.
$$
Therefore the sum of ${\cal N}$ `small' centered matrices has a Wigner spectrum in the large ${\cal N}$ limit, with computable corrections.

The next example is to consider empirical correlation in the case where the true correlation matrix is the 
identity: $\mathbf{C}=\mathbf{I}$. Then, $\mathbf{E}$ is by definition the sum of rank one matrices 
$\delta E^t_{ij}=(r_i^t r_j^t)/T$, where $r_i^t$ are independent, unit variance random variables. Hence,
$\delta \mathbf{E}^t$ has one eigenvalue equal to $q$ (for large $N$) associated with direction ${\bf r}^t$, and $N-1$ zero eigenvalues corresponding to the hyperplane perpendicular to ${\bf r}^t$. The different $\delta \mathbf{E}^t$ are therefore mutually free and one can use the R-transform trick. Since:
\be
\delta G^t(z)=\frac{1}{N}\left(\frac{1}{z-q}+\frac{N-1}{z}\right)
\ee
Inverting $\delta G(z)$ to first order in $1/N$, the elementary Blue transform reads: 
\be
\delta B(z)=\frac{1}{z}+\frac{q}{N(1-qz)} \longrightarrow \delta R(z)=\frac{q}{N(1-qz)}.
\ee
Using the addition of R-transforms, one then deduces:
\be\label{MPG}
B_{E}(z)=\frac{1}{z}+\frac{1}{(1-qz)} \longrightarrow
G_E(z)=\frac{(z+q-1)-\sqrt{(z+q-1)^2-4zq}}{2 z q},
\ee
which reproduces the well known Mar\v{c}enko \& Pastur result for the density of eigenvalues (for $q < 1$) \cite{MP}:
\be
\rho_E(\lambda) = \frac{\sqrt{4\lambda q - (\lambda+q-1)^2}}{2\pi \lambda q}, \quad \lambda \in [(1-\sqrt{q})^2,
(1+\sqrt{q})^2].
\ee
This distribution is plotted in Fig. 1 for $Q=1/q=3.45$. The remarkable feature of this result is that there should be {\it no} eigenvalue outside the interval $[(1-\sqrt{q})^2,
(1+\sqrt{q})^2]$ when $N \to \infty$.
One can check that $\rho_E(\lambda)$ converges towards $\delta(\lambda-1)$ when $q=1/Q \to 0$, or $T \gg N$. When $q > 1$,
we know that some zero eigenvalues necessarily appear in the spectrum, which then reads:
\be
\rho_E(\lambda) = (1-Q) \delta(\lambda) + \frac{\sqrt{4\lambda Q - (\lambda+Q-1)^2}}{2\pi \lambda}
\ee
Using $G_E(z)$, it is straightforward to show that $(1/N)\Tr {\bf E}^{-1}=-G_E(0)$ is given by $(1-q)^{-1}$ for $q < 1$.
This was alluded to in Sect. \ref{sect2.3} above. The Mar\v{c}enko-Pastur is important because of its large degree of universality: as for the Wigner semi-circle, 
its holds whenever the random variables $r_i^t$ are IID with a finite second moment (but see Sect. \ref{sect3.4} below for other `universality classes').
\footnote{A stronger statement, on which we will return below, is that provided the $r_i^t$ have a finite {\it fourth} moment, the largest eigenvalue
of the empirical correlation matrix ${\bf E}$ tends to the upper edge of the Mar\v{c}enko-Pastur distribution, $\lambda_+=(1+\sqrt{q})^2$.}

\begin{figure}
\begin{center}
\psfig{file=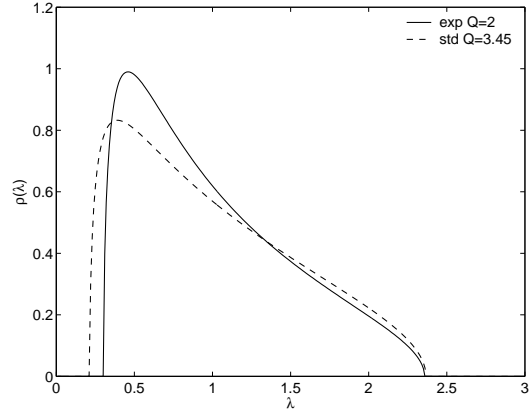,width=7cm} 
\end{center}
\caption{Mar\v{c}enko \& Pastur spectrum for $Q=T/N=3.45$ (dotted line) compared to the spectrum of the exponentially weighted moving average 
correlation random matrix with $q\equiv N\epsilon=1/2$ (plain line).
}
\label{Fig1}
\end{figure}

Consider now the case where the empirical matrix is computed using an exponentially 
weighted moving average (still with $\mathbf{C}=\mathbf{I}$). Such an estimate is standard practice in finance. More precisely:
\be\label{EMA}
E_{ij}=\epsilon\sum_{t'=-\infty}^{t-1}(1-\epsilon)^{t-t'} r_i^{t'} r_j^{t'} 
\ee
with $\epsilon<1$. Now, as an ensemble $E_{ij}$ satisfies $E_{ij}=(1-\epsilon) E_{ij} + 
\epsilon r_i^{0} r_j^{0}$. We again invert the resolvent of $\mathbf{E}_0$ to find the elementary Blue transform,
\be
B_0(z)=\frac{1}{z}+R_0(z)\qquad\mbox{{with}}\qquad R_0(z)=\frac{q}{N(1-qx)}
\ee
where now $q=N\epsilon$. Using again Eq. (\ref{Rscaling}), we then find for $R(z)$, to first order in $1/N$:
\be
R(z)+zR'(z)+\frac{q}{1-qz}=0 \longrightarrow R(z)=-\frac{\log(1-qz)}{qz}.
\ee
Going back to the resolvent to find the density, we finally get \cite{KondorPotters}:
\be
\rho(\lambda) = \frac{1}{\pi} \Im G(\lambda)\quad\mbox{{ where $G(\lambda)$ solves }}
\quad\lambda q G=q-\log(1-qG)
\ee
This solution is compared to the standard Wishart distribution in Fig 1.

A nice property of the Blue functions is that they can be used to find the edges
of the eigenvalue spectrum ($\lambda_\pm$). One has:\cite{zee}
\be
\lambda_\pm =B(z_\pm)\qquad\mbox{where}\qquad B'(z_\pm)=0
\ee
In the case at hand, by evaluating $B(z)$ when $B'(z)=0$ we can write 
directly an equation whose solutions are the spectrum edges ($\lambda_\pm$)
\be
\lambda_\pm=\log(\lambda_\pm)+q+1
\ee
When $q$ is zero, the spectrum is again $\delta(\lambda-1)$ as expected. 
But as the noise increases (or the characteristic time decreases) the lower edge approach zero very quickly as $\lambda_-\sim\exp(-1/Q)$. 
Although there are no exact zero eigenvalues for these matrices, the smallest eigenvalue is exponentially close to zero when $Q \to 0$, i.e. $N \gg T$.

\subsection{More applications}
\label{sect3.4}
\subsubsection{The case of an arbitrary true correlation matrix}

In general, the random variables under consideration are described by `true' correlation matrix $\mathbf{C}$ with 
some non trivial structure, different from the identity matrix ${\bf 1}$. Interestingly, the Mar\v{c}enko-Pastur result for the spectrum of the empirical matrix ${\bf E}$ can be extended to a rather general $\mathbf{C}$, and opens the way to 
characterize the true spectrum $\rho_{C}$ even with partial information $Q=T/N < \infty$. However, for a general
$\bf C$, the different projectors $r_i^t r_j^t$ cannot be assumed to define uncorrelated directions for different $t$,
even if the random variables $r_i^t$ are uncorrelated in time and the above trick based on R-transforms cannot be used.
However, assuming that the $r_i^t$ are Gaussian, the empirical matrix ${\bf E}$ can always be written as 
${\bf C}^{1/2}\hat{\bf X}[{\bf C}^{1/2}\hat{\bf X}]^{{{T}}}$, where $\hat {\bf X}$ is an $N \times T$ rectangular matrix of uncorrelated, unit variance Gaussian random variables. But since the eigenvalues of 
${\bf C}^{1/2}\hat{\bf X}[{\bf C}^{1/2}\hat{\bf X}]^{{{T}}}$ are the same as those of ${\bf C}\hat{\bf X}\hat{\bf X}^{{{T}}}$, we can use the S-transform trick mentioned above, with ${\bf A}={\bf C}$ and ${\bf B}=
\hat{\bf X}\hat{\bf X}^{{{T}}}$ mutually free, and where the spectrum of ${\bf B}$ is by construction given 
by the Mar\v{c}enko-Pastur law. This allows one to write down the following self-consistent for the resolvent of ${\bf E}$:
\footnote{Other techniques, such as the Replica method, or the summation of planar diagrams, can also be used
to obtain this result.}\cite{baisilverstein,Burda1}
\be\label{burda_basic}
G_E(z)= \int d\lambda\, \rho_C(\lambda) \frac{1}{z-\lambda (1-q+qz G_E(z))},
\ee
a result that in fact already appears in the original Mar\v{c}enko-Pastur paper! One can check that if $\rho_C(\lambda)=\delta(\lambda-1)$, one recovers the
result given by Eq. (\ref{MPG}). Equivalently, the above relation can be written as:
\be\label{burda}
zG_E(z) = Z G_C(Z) \qquad\mbox{where} \qquad Z = \frac{z}{1+q(zG_E(z) -1)},
\ee 
which is convenient for numerical evaluation \cite{Burda1}. From these equations, one can evaluate $-G_E(0)=  \Tr {\bf E}^{-1}$,
which is found to be equal to $\Tr {\bf C}^{-1}/(1-q)$, as we mentioned in the introduction, and used to derive Eq. (\ref{risks}) above.

Note that while the mapping between the true spectrum $\rho_C$
and the empirical spectrum $\rho_E$ is numerically stable, the inverse mapping is unstable, a little bit
like the inversion of a Laplace transform. In order to reconstruct the spectrum of ${\bf C}$ from that of 
${\bf E}$ one should therefore use a parametric ansatz of $\rho_C$ to fit the observed $\rho_E$, and not try
to invert directly the above mapping (for more on this, see \cite{Burda2,NEK}).

Note also that the above result does {\it not} apply when $\mathbf{C}$ has isolated eigenvalues, and
only describes continuous parts of the spectrum. For example, if one considers a matrix $\mathbf{C}$
with one large eigenvalue that is separated from the `Wishart sea', the statistics of this isolated
eigenvalue has recently been shown to be Gaussian \cite{BBP} (see also below), with a width $\sim T^{-1/2}$, 
much smaller than the uncertainty on the bulk eigenvalues ($\sim q^{1/2}$). A naive application of Eq. (\ref{burda}),
on the other hand, would give birth to a `mini-Wishart' distribution around the top eigenvalue. This
would be the exact result only if the top eigenvalue of $C$ had a degeneracy proportional to $N$. 

\subsubsection{The Student ensemble case}

Suppose now that the $r_i^t$ are chosen according to the Student multivariate distribution described in Sect. \ref{sect2.2} above.
Since in this case $r_i^t = \sigma_t \xi_i^t$, the empirical correlation matrix can be written as:
\be
E_{ij} = \frac1T \sum_t \sigma_t^2 \xi_i^t \xi_j^t, \qquad \langle \xi_i \xi_j \rangle \equiv \hat C_{ij}
\ee
In the case where $\hat{\bf C}={\bf 1}$, this can again be seen as a sum of mutually free projectors, and one can use the
R-transform trick. This allows one to recover the following equation for the resolvent of ${\bf E}$, first obtained in the Mar\v{c}enko-Pastur paper and exact in the large $N,T$ limit:
\begin{eqnarray}\label{doseq1}
\lambda&=&\frac{G_R}{G_R^2+\pi^2 \rho_E^2}+\int ds P(s) \frac{\mu (s-q\mu G_R)}{(s-q\mu G_R)^2+\pi^2\rho_E^2}\\
0&=&\rho\left(-\frac{1}{G_R^2\pi^2\rho_E^2}+\int ds P(s) \frac{q\mu^2}{(s-q\mu G_R)^2+\pi^2\rho_E^2}
\right),\label{doseq2}
\end{eqnarray}
where $G_R$ is the real part of the resolvent, and $P(s)=s^{\mu/2-1} e^{-s}/\Gamma(\mu/2)$ is the distribution
of $s=\mu/\sigma^2$ in the case of a Student distribution; however, the above result holds for other distributions
of $\sigma$ as well, corresponding to the class of ``elliptic'' multivariate distributions. The salient results are \cite{BiroliStudent}: (i) there is no longer any upper edge of the spectrum: 
$\rho_E(\lambda) \sim \lambda^{-1-\mu/2}$ when $\lambda \to \infty$; (ii) but there is a lower edge to the 
spectrum for all $\mu$. The case $\hat{\bf C} \neq {\bf 1}$ can also be treated using S-transforms. 

Instead of the usual (Pearson) estimate of the correlation matrix, one could use the maximum likelihood procedure, 
Eq. (\ref{maxlstudent}) above. Surprisingly at first sight, the corresponding spectrum $\rho_{ML}(\lambda)$ is
then completely different \cite{BiroliStudent}, and is given by the standard Mar\v{c}enko-Pastur result! The intuitive reason is that the
maximum likelihood estimator Eq. (\ref{maxlstudent}) effectively renormalizes the returns by the daily volatility
$\sigma_t$ when $\sigma_t$ is large. Therefore, all the anomalies brought about by `heavy days' (i.e.
$\sigma_t \gg \sigma_0$) disappear.

Finally, we should mention that another Student Random-Matrix ensemble has been considered in the recent literature, where
instead of having a time dependent volatility $\sigma_t$, it is the global volatility $\sigma$ that is random, and distributed 
according to Eq. (\ref{defstudent}) \cite{Bohigas,BurdaStudent,Vivo}. The density of states is then simply obtained by averaging over 
Mar\v{c}enko-Pastur distributions of varying width. Note however that in this case the density of states is {\it not} self-averaging:
each matrix realization in this ensemble will lead to a Mar\v{c}enko-Pastur spectrum, albeit with a random width.

\subsection{Random SVD}
\label{sect3.5}
As we mentioned in Sect. \ref{SVD_intro}, it is often interesting to consider non-symmetrical, or even rectangular
correlation matrices, between $N$ `input' variables $X$ and $M$ `output' variables $Y$. The empirical correlation
matrix using $T$-long times series is defined by Eq. (\ref{empiricalSVD}). What can be said about the singular
value spectrum of ${\cal E}$ in the special limit $N,M,T \to \infty$, with $n=N/T$ and $m=M/T$ fixed? Whereas the
natural null hypothesis for correlation matrices is ${\bf C}={\bf 1}$, that leads to the Mar\v{c}enko-Pastur density, 
the null hypothesis for cross-correlations between {\it a priori} unrelated sets of input and output variables is
${\cal C}={\bf 0}$. However, in the general case, input and output variables can very well be correlated between
themselves, for example if one chooses redundant input variables. In order to establish a universal result, 
one should therefore consider the {\it exact} normalized principal components for the sample variables $X$'s and $Y$'s: 
\be
\hat X_\alpha^t = \frac{1}{\sqrt{\lambda_\alpha}} \sum_i V_{\alpha,i} X_i^t;.
\ee
and similarly for the $\hat Y_a^t$. The $\lambda_\alpha$ and the $V_{\alpha, i}$ are the eigenvalues
and eigenvectors of the sample correlation matrix ${\bf E}_X$ (or, respectively ${\bf E}_Y$). We now define the normalized
$M \times N$ cross-correlation matrix as $\hat {\cal E} = \hat Y \hat X^{{{T}}}$. One can then use the following tricks \cite{RSVD}:
\begin{itemize}
\item The non zero eigenvalues of $\hat{\cal E}^{{{T}}} \hat{\cal E}$ are the same as those of $\hat X^{{{T}}}\hat X 
\hat Y^{{{T}}} \hat Y$
\item ${\bf A}=\hat X^{{{T}}} \hat X$ and ${\bf B}=\hat Y^{{{T}}} \hat Y$ are two mutually free $T \times T$
matrices, with $N$ ($M$) eigenvalues exactly equal to $1$ (due to the very construction of $\hat X$ and 
$\hat Y$), and $(T-N)^+$ ($(T-M)^+$) equal to $0$.
\item The S-transforms are multiplicative, allowing one to obtain the spectrum of ${\bf AB}$.
\end{itemize}

Due to the simplicity of the spectra of ${\bf A}$ and ${\bf B}$, the calculation of S-transforms is particularly 
easy \cite{RSVD}. The final result for the density of singular values (i.e, the square-root of the eigenvalues of ${\bf AB}$) reads (see 
\cite{Wachter} for an early derivation of this result, see also \cite{IMJ2}):
\be\label{resultSVD}
\rho(c) = \max(1-n,1-m) \delta(c) + \max(m+n-1,0) \delta(c-1) + \Re \frac{\sqrt{(c^2-\gamma_-)(\gamma_+-c^2)}}{\pi c(1 - c^2)},
\ee
where $n=N/T$, $m=M/T$ and $\gamma_\pm$ are given by: 
\be
\gamma_{\pm} = n + m - 2mn \pm 2\sqrt{mn(1 - n)(1 - m)}, \quad 0 \leq \gamma_{\pm} \leq 1
\ee
The allowed $c$'s are all between $0$ and $1$, as they
should since these singular values can be interpreted as correlation coefficients. In the limit $T \to \infty$ at fixed $N$, $M$, all singular values collapse to zero, as they should
since there is no true correlations between $X$ and $Y$; the allowed band in the limit $n,m \to 0$ becomes:
\be
c \in \left[\frac{|m-n|}{\sqrt{m}+\sqrt{n}},\frac{m-n}{\sqrt{m}-\sqrt{n}}
\right],
\ee
showing that for fixed $N,M$, the order of magnitude of allowed singular values decays as $T^{-1/2}$.

Note that one could have considered a different benchmark ensemble, where one considers two independent vector time series $X$ and $Y$ with true correlation matrices 
$C_X$ and $C_Y$ equal to ${\bf 1}$. The direct SVD spectrum in that case can also be computed as the S-convolution of two Mar\v{c}enko-Pastur distributions with parameters $m$ and $n$
\cite{RSVD}, This alternative benchmark is however not well suited in practice, since it mixes up the possibly non trivial correlation structure of the input variables 
and of the output variables themselves with the {\it cross}-correlations between these variables.

As an example of applications to economic time series, we have studied in \cite{RSVD} the cross correlations between 
76 different macroeconomic indicators (industrial production, retail sales, new orders and inventory indices of all economic activity sectors available, etc.) 
and 34 indicators of inflation, the Composite Price Indices (CPIs), concerning different sectors of activity during the period 
June 1983-July 2005, corresponding to 265 observations of monthly data. The result is that only one, or perhaps two singular values emerge from the above ``noise band''.
From an econometric point of view, this is somewhat disappointing: there seems to be very little exploitable signal in spite of the quantity of available observations. 

\subsection{A Note on ``L\'evy'' (or heavy tailed) matrices  \cite{Burdabook}}
\label{sect3.6}
All the above results for the bulk part of the spectrum of random matrices are to a large extent universal 
with respect to the distribution of the matrix elements. Although many of these results are easy to obtain
assuming that the random variables involved in their construction are Gaussian, this is not a crucial assumption.
For example, the Wigner semi-circle distribution holds for any large symmetric matrices made up of IID elements, 
provided these have a finite second moment. 

The results are however expected to change when the tail of the distribution of these elements are so heavy that the
second moment diverges, corresponding to a tail index $\mu$ less than $2$. The generalization of the Wigner distribution in that case 
(called L\'evy matrices, because the corresponding ensemble is stable under addition) was worked out in \cite{CB} using heuristic methods  \cite{Burdabook}.
Their result on $\rho(\lambda)$ was recently rigorously proven in \cite{BAG}. The remarkable feature is that the support of the 
spectrum becomes unbounded; actually $\rho(\lambda)$ decays for large $\lambda$ with the exact same tail as that of the distribution of 
individual elements. 

It is worth noticing that although L\'evy matrices are by construction stable under addition, two such L\'evy
matrices are {\it not} mutually free. The problem comes in particular from the large eigenvalues just mentioned; the corresponding eigenvectors 
are close to one of the canonical basis vector. Therefore one cannot assume that the eigenbasis differ by a random rotation. 
A different ensemble can however be constructed, where each L\'evy matrix
is randomly rotated before being summed (see \cite{Burdabook}). 
In this case, freeness is imposed by hand and R-transforms are additive. The corresponding fixed point generalizing $R(z)=z$ in the Wigner case is then $R(z)=z^{\mu-1}$. The eigenvalue spectrum is however different
from the one obtained in \cite{CB,BAG}, although the asymptotic tails are the same: $\rho(\lambda) \propto \lambda^{-1-\mu}$.

Finally, the generalization of the Mar\v{c}enko-Pastur result for heavy tailed matrices is also a very recent achievement \cite{DG}.
Again, the spectrum loses both its upper and lower sharp edges for all finite values of $Q=T/N$ as soon as the variance of 
the random variables $r_i^t$ diverges, i.e. when $\mu < 2$. Note that the resulting spectrum is distinct
from the Student ensemble result obtained above, the latter is different from Mar\v{c}enko-Pastur for all $\mu < + \infty$. 
However, when $\mu < 2$, they both share the same power-law tail which is now: $\rho(\lambda) \propto \lambda^{-1-\mu/2}$.

\section{Random Matrix Theory: The Edges}

\subsection{The Tracy-Widom region}
\label{sect4.1}
As we alluded to several times, the practical usefulness of the above predictions for the eigenvalue spectra of
random matrices is (i) their universality with respect to the distribution of the underlying random variables and
(ii) the appearance of sharp edges in the spectrum, meaning that the existence of eigenvalues lying outside the 
allowed band is a strong indication against several null hypothesis benchmarks. 

However, the above statements are only true in the asymptotic, $N,T \to \infty$ limit. For large but finite $N$ one expects that the probability to find an eigenvalue is very small but finite. 
The width of the transition region, and
the tail of the density of states was understood a while ago, culminating in the beautiful results by Tracy \& Widom on
the distribution of the {\it largest} eigenvalue of a random matrix. There is now a huge literature on this topic 
(see e.g. \cite{IMJ2,Johansson,BBP,Peche,Majumdar}) that we will not attempt to cover here in details. We will only extract
a few interesting results for applications.

The behaviour of the width of the transition region can be understood using a simple heuristic argument. Suppose that
the $N=\infty$ density goes to zero near the upper edge $\lambda_+$ as $(\lambda_+-\lambda)^\theta$ (generically, 
$\theta=1/2$ as is the case for the Wigner and the Mar\v{c}enko-Pastur distributions). For finite $N$, one expects
not to be able to resolve the density when the probability to observe an eigenvalue is smaller than $1/N$. This criterion reads:
\be
(\lambda_+-\lambda^*(N))^{\theta+1} \propto \frac1N \to \Delta \lambda^* \sim N^{-\frac{1}{1+\theta}},
\ee
or a transition region that goes to zero as $N^{-2/3}$ in the generic case. More precisely, for Gaussian ensembles, the average 
density of states at a distance $\sim N^{-2/3}$ from the edge behaves as:
\be
\rho_N(\lambda \approx \lambda_+) = N^{-1/3} \Phi\left[N^{2/3}(\lambda-\lambda_+)\right],
\ee
with $\Phi(x \to -\infty) \propto \sqrt{-x}$ as to recover the asymptotic density of states, and $\ln \Phi(x \to +\infty) \propto x^{3/2}$,
showing that the probability to find an eigenvalue outside of the allowed band decays exponentially with $N$ and super exponentially with 
the distance to the edge. 

A more precise result concerns the distance between the largest eigenvalue $\lambda_{\max}$ of a random matrix and the upper edge of the spectrum
$\lambda_+$. The Tracy-Widom result is that for a large class of $N \times N$ matrices (e.g. symmetric random matrices with IID elements with a finite fourth
moment, or empirical correlation matrices of IID random variables with a finite fourth moment), the rescaled distribution of $\lambda_{\max}-\lambda^*$
converges towards the Tracy-Widom distribution, usually noted $F_1$:
\be
\mbox{Prob}\left(\lambda_{\max} \leq \lambda_+ + \gamma N^{-2/3} u\right)=F_1(u),
\ee
where $\gamma$ is a constant that depends on the problem. For example, for the Wigner problem, $\lambda_+=2$ and $\gamma=1$; whereas
for the Mar\v{c}enko-Pastur problem, $\lambda_+=(1+\sqrt{q})^2$ and $\gamma=\sqrt{q}\lambda_+^{2/3}$.

Everything is known about the Tracy-Widom density $f_1(u)=F_1'(u)$, in particular its left and right far tails:
\be
\ln f_1(u) \propto -u^{3/2}, \quad (u \to +\infty); \qquad \ln f_1(u) \propto -|u|^{3}, \quad (u \to -\infty);
\ee
Not surprisingly, the right tail is the same as that of the density of states $\Phi$. The left tail is much thinner:
pushing the largest eigenvalue inside the allowed band implies compressing the whole Coulomb-Dyson gas of charges, which
is difficult. Using this analogy, the large deviation regime of the Tracy-Widom problem (i.e. for $\lambda_{\max} - \lambda_+= O(1)$)
can be obtained \cite{Dean-Majumdar}.

Note that the distribution of the smallest eigenvalue $\lambda_{\min}$ around the lower edge $\lambda_-$ is also
Tracy-Widom, except in the particular case of Mar\v{c}enko-Pastur matrices with $Q=1$. In this case, $\lambda_-=0$
which is a `hard' edge since all eigenvalues of the empirical matrix must be non-negative. This special case is treated
in, e.g. \cite{Peche}.

Finally, the distance of the largest singular value from the edge of the random SVD spectrum, Eq. (\ref{resultSVD}) above, is also 
governed by a Tracy-Widom distribution, with parameters discussed in details in \cite{IMJ3}.

\subsection{The case with large, isolated eigenvalues and condensation transition}
\label{sect4.2}
The Wigner and Mar\v{c}enko-Pastur ensembles are in some sense maximally random: no prior information on the structure
of the matrices is assumed. For applications, however, this is not necessarily a good starting point. In the example of
stock markets, it is intuitive that all stocks are sensitive to global news about the economy, for example. This means 
that there is at least one common factor to all stocks, or else that the correlation coefficient averaged over all
pairs of stocks, is positive. A more reasonable null-hypothesis is that the true correlation matrix is: $C_{ii}=1$, 
$C_{ij}=\overline{\rho}$, $\forall i \neq j$. This essentially amounts to adding to the empirical correlation matrix a rank one
perturbation matrix with one large eigenvalue $N\overline{\rho}$, and $N-1$ zero eigenvalues. 
When $N\rho \gg 1$, the empirical correlation
matrix will obviously also have a large eigenvalue close to $N \rho$, very far above the Mar\v{c}enko-Pastur
upper edge $\lambda_+$. What happens when $N \overline{\rho}$ is not very large compared to unity?

This problem was solved in great details by Baik, Ben Arous and P\'ech\'e \cite{BBP}, who considered the more general case where
the true correlation matrix has $k$ special eigenvalues, called ``spikes''. A similar problem arises when one considers Wigner matrices, 
to which one adds a perturbation matrix of rank $k$. For example, if the random elements $H_{ij}$ have a non zero
mean $\overline{h}$, the problem is clearly of that form: the perturbation has one non zero eigenvalue $N \overline{h}$,
and $N-1$ zero eigenvalues. As we discuss now using free random matrix techniques, this problem has a sharp phase 
transition between a regime where this rank one perturbation is weak and is ``dissolved'' in the Wigner 
sea, and a regime where this perturbation is strong enough to escape from the Wigner sea. This transition corresponds 
to a ``condensation'' of the eigenvector corresponding to the largest eigenvalue onto the eigenvalue of the rank one
perturbation. 

Let us be more precise using R-transform techniques for the Wigner problem. Assume that the non zero eigenvalue of the 
rank one perturbation is $\Lambda$, with a corresponding eigenvector ${\vec e}_1=(1,0,\dots,0)$. The resolvent $G_\Lambda$ and
the Blue function $B_\Lambda$ of this perturbation is:
\be
G_\Lambda(z)=\frac{N-1}{Nz} + \frac{1}{N} \frac{1}{z - \Lambda} \to B_\Lambda(z) \approx \frac1z + 
\frac{1}{N} \frac{\Lambda}{1 - \Lambda z}
\ee
Such a perturbation is free with respect to Wigner matrices. The R-transform of the sum is therefore given by:
\be
R_{H+\Lambda}=z+\frac{1}{N} \frac{\Lambda}{1 - \Lambda z} \to z \approx G + \frac1G + \frac{1}{N} 
\frac{\Lambda}{1 - \Lambda G}
\ee
which allows to compute the corrected resolvent $G$. The correction term is of order $1/N$, and one can substitute $G$ by the 
Wigner resolvent $G_W$ to first order. This correction can only survive in the
large $N$ limit if $\Lambda \times G_W(z)=1$ has a non trivial solution, such that the divergence compensates the $1/N$ factor. 
The corresponding value of $z$ then defines an isolated eigenvalue.
This criterion leads to \cite{Feral,BiroliTails}:
\be
z = \lambda_{\max} = \Lambda + \frac1\Lambda \quad (\Lambda>1); \qquad \lambda_{\max} = 2 \quad (\Lambda \leq 1)
\ee

Therefore, the largest eigenvalue pops out of the Wigner sea precisely when $\Lambda = 1$. The statistics of the largest
eigenvalue $\lambda_{\max}$ is still Tracy-Widom whenever $\Lambda < 1$, but becomes Gaussian, of width $N^{-1/2}$ 
(and not $N^{-2/3}$) when $\Lambda > 1$. The case $\Lambda=1$ is special and is treated in \cite{BBP}. Using 
simple perturbation theory, one can also compute the overlap between the largest eigenvector $\vec V_{\max}$ 
and ${\vec e}_1$ \cite{BiroliTails}:
\be
(\vec V_{\max} \cdot {\vec e}_1)^2 = 1 - \Lambda^{-2}, \quad (\Lambda>1),
\ee
showing that indeed, the coherence between the largest eigenvector and the perturbation becomes progressively lost
when $\Lambda \to 1^+$.

A similar phenomenon takes place for correlation matrices. For a rank one perturbation of the type described above, with 
an eigenvalue $\Lambda=N\rho$, the criterion for expelling an isolated eigenvalue from the Mar\v{c}enko-Pastur sea now
reads \cite{BBP}:
\be
\lambda_{\max} = \Lambda + \frac{\Lambda q}{\Lambda-1} \quad (\Lambda>1+\sqrt{q}); \qquad \lambda_{\max} = (1+\sqrt{q})^2 
\quad (\Lambda \leq 1+\sqrt{q})
\ee
Note that in the limit $\Lambda \to  \infty$, $\lambda_{\max} \approx \Lambda + q + O(\Lambda^{-1})$.
For rank $k$ perturbation, all eigenvalues such that $\Lambda_r>1+\sqrt{q}$, $1 \leq r \leq k$ 
will end up isolated above the Mar\v{c}enko-Pastur sea, all others disappear below $\lambda_+$. 
All these isolated eigenvalues have Gaussian fluctuations of order $T^{-1/2}$ (see also Sect. \ref{sect4.4} below). For more 
details about these results, see \cite{BBP}.

\subsection{The largest eigenvalue problem for heavy tailed matrices}
\label{sect4.3}
The Tracy-Widom result for the largest eigenvalue was first shown for the Gaussian Orthogonal ensemble, but it was
soon understood that the result is more general. In fact, if the matrix elements are IID with a finite fourth moment,
the largest eigenvalue statistics is {\it asymptotically} governed by the Tracy-Widom mechanism. Let us give a few
heuristic arguments for this \cite{BiroliTails}. Suppose the matrix elements are IID with a power-law distribution:
\be
P(H) \sim_{|H|\to \infty} \frac{A^{\mu}}{|H|^{1+\mu}}\, \mbox{ with } \, A \sim O(1/\sqrt{N}).
\ee
and $\mu > 2$, such that the asymptotic eigenvalue spectrum is the Wigner semi-circle with $\lambda_{\pm}=\pm 2$.
The largest element $H_{\max}$ of the matrix (out of $N^2/2$) is therefore of order  $N^{2/\mu-1/2}$
and distributed with a Fr\'echet law. From the results of the previous subsection, one can therefore expect that:
\begin{itemize}
\item If $\mu>4$: $H_{\max} \ll 1$, and one recover Tracy-Widom.
\item If $2<\mu<4$: $H_{\max} \gg 1$, $\lambda_{\max} \approx H_{\max} \propto N^{\frac{2}{\mu} - \frac{1}{2}}$, 
with a Fr\'echet distribution. Note that although $\lambda_{\max} \to \infty$ when $N \to \infty$, 
the density itself goes to zero when $\lambda > 2$ in the same limit.
\item If $\mu=4$: $H_{\max} \sim O(1)$, $\lambda_{\max} = 2$ or $\lambda_{\max} = H_{\max}+1/H_{\max}$, corresponding
to a non-universal distribution for $\lambda_{\max}$ with a $\delta$-peak at $2$ and a transformed Fr\'echet distribution
for $\lambda_{\max} > 2$.
\end{itemize}

Although the above results are expected to hold for $N \to \infty$ (a rigorous proof can be found in \cite{BAP}), 
one should note that there are very strong finite size corrections. In particular, although for $\mu > 4$ the 
asymptotic limit is Tracy-Widom, for any finite $N$ the distribution of the largest eigenvalue has power-law tails
that completely mask the Tracy-Widom distribution -- see \cite{BiroliTails}. Similarly, the convergence towards the Fr\'echet 
distribution for $\mu < 4$ is also very slow.

\subsection{Dynamics of the top eigenvector -- theory}
\label{sect4.4}
As mentioned above and discussed in fuller details in the next section, 
financial covariance matrices are such that a few large eigenvalues are 
well separated from the `bulk', where all other eigenvalues reside. 
We have indeed seen that if stocks tend to be correlated on average, a large
eigenvalue $\lambda_{\max} \approx N \overline{\rho}$ will emerge. The associated
eigenvector is the so-called `market mode': in a first approximation, all stocks move together, up or down. 

A natural question, of great importance for portfolio management
is whether $\lambda_{\max}$ and the corresponding $\vec V_{\max}$ are stable in time. Of course, 
the largest eigenvalue and eigenvector of the
empirical correlation matrix are affected by measurement noise. Can one make 
predictions about the fluctuations of both the largest eigenvalue and the corresponding eigenvector induced by 
measurement noise? This would help separating a true evolution in time of the average stock correlation 
and of the market exposure of each stock from one simply related to measurement noise. Such a 
decomposition seems indeed possible in the limit where $\lambda_{\max} \gg \lambda_\alpha$. 

Suppose that the true covariance matrix $\bf C$ 
is time independent with one large eigenvalue $\Lambda_1$ associated to the normalized eigenvector $\vec V_1$. 
Assuming that the covariance matrix ${\bf E}_t$ is measured through an exponential moving average of the returns, Eq. (\ref{EMA}),
with an averaging time $1/\epsilon$, one can write down, in the limit $\epsilon \to 0$ and for Gaussian returns, two Ornstein-Uhlenbeck like equations
for the largest eigenvalue of ${\bf E}_t$, $\lambda_{1t}$, and for its associated eigenvector $\vec v_{1t}$ \cite{Krakow1}. The angle $\theta$ between
$\vec v_{1t}$ and $\vec V_1$ reaches a stationary distribution given by:
\be
P(\theta) = {\cal N} \left[\frac{1 + \cos 2\theta (1 - \frac{\Lambda_b}{\Lambda_1})}
{1 - \cos 2\theta (1 - \frac{\Lambda_1}{\Lambda_0})}\right]^{1/4\epsilon}
\ee
where $\Lambda_b$ is the average value of the bulk eigenvalues of ${\bf C}$, assumed to be $\ll \Lambda_1$.
As expected, this distribution is invariant when $\theta \to \pi - \theta$, since $-\vec V_1$ is also a top eigenvector.
In the limit $\Lambda_b \ll \Lambda_1$, one sees that the distribution becomes peaked around $\theta=0$ and $\pi$.
For small $\theta$, the distribution is Gaussian, with $\langle \cos^2 \theta \rangle 
\approx 1 - \epsilon {\Lambda_b}/2{\Lambda_1}$. The angle $\theta$ is less 
and less fluctuating as $\epsilon \to 0$ (as expected) but also as $\Lambda_b/\Lambda_1 \to 0$: a large separation of
eigenvalues leads to a well determined top eigenvector. In this limit, the
distribution of $\lambda_1$ also becomes Gaussian (as expected from general results \cite{BBP}) and one finds, to leading order:
\be
\langle \lambda_1 \rangle \approx \Lambda_1 - \epsilon {\Lambda_b}/2; \qquad 
\langle(\delta \lambda_1)^2\rangle \approx \Lambda_1^2 \epsilon.
\ee
In the limit of large averaging time and one large top eigenvalue (a situation 
approximately realized for financial markets), the deviation from the true top eigenvalue 
$\delta \lambda_1$ and the deviation angle $\theta$ are independent Gaussian variables.
One can compute the variogram of the top eigenvalue as:
\be\label{varval}
\langle[\lambda_{1,t+\tau}-\lambda_{1,t}]^2\rangle
=2 \Lambda_1^2 \epsilon \left(1 - \exp(-\epsilon \tau)\right).
\ee
One can also work out the average overlap of the top eigenvector
with itself as a function of time lag, leading to:
\be\label{varvec}
\langle (\vec v_{1t}-\vec v_{1t+\tau})^2 \rangle = 2 - 2\langle \cos(\theta_t - \theta_{t+\tau}) \rangle \approx 
2\epsilon \frac{\Lambda_b}{\Lambda_1} (1 - \exp(-\epsilon \tau)).
\ee
These results assume that ${\bf C}$ is time independent. Any significant deviation from the above laws would indicate a genuine 
evolution of the market structure. We will come back to this point in section \ref{sect5.3}.

\section{Applications: Cleaning correlation matrices}

\subsection{Empirical eigenvalue distributions}
\label{sect5.1}

Having now all the necessary theoretical tools in hand, we turn to the analysis of empirical correlation matrices of stock returns. Many 
such studies, comparing the empirical spectrum with RMT predictions, have been published in the literature. Here, we perform this analysis once more, on an 
extended data set, with the objective of comparing precisely different cleaning schemes for risk control purposes (see next subsection, \ref{sect5.2}).

We study the set of U.S. stocks between July, 1993 and April, 2008 (3700 trading days). We consider 26 samples obtained by sequentially sliding a window of $T=1000$ days 
by 100 days. For each period, we look at the empirical correlation matrix of the $N=500$ most liquid stocks during that period. The quality factor is therefore $Q=T/N=2$.
The eigenvalue spectrum shown in Fig. 2 is an average over the 26 sample eigenvalue distributions, where we have removed the market mode and rescaled the eigenvalues
such that $\int {\rm d}\lambda \rho_E(\lambda)=1$ for each sample. The largest eigenvalue contributes on average to $21 \%$ of the total trace. 

We compare in Fig. 2 the empirical spectrum with the Mar\v{c}enko-Pastur prediction for $Q=1/q=2$. It is clear that several eigenvalues leak out of the Mar\v{c}enko-Pastur
band, even after taking into account the Tracy-Widom tail, which have a width given by $\sqrt{q}\lambda_+^{2/3}/N^{2/3} \approx 0.02$, very small in the present case. The 
eigenvectors corresponding to these eigenvalues show significant structure, that correspond to identifiable economic sectors. Even after accounting for these large 
eigenvalues, the Mar\v{c}enko-Pastur prediction is not very good, suggesting that the prior for the underlying correlation matrix ${\bf C}$ may carry more 
structure than just a handful of eigenvalue ``spikes'' on top of the identity matrix \cite{Burda2,Malevergne}. An alternative simple prior for the spectrum of 
${\bf C}$ is a power-law distribution, corresponding to the coexistence of large sectors and smaller sectors of activity:
\be\label{powerlaw}
\rho_C(\lambda) = \frac{\mu A}{(\lambda - \lambda_0)^{1+\mu}} \Theta(\lambda - \lambda_{\min}),
\ee
with $A$ and $\lambda_0$ related to $\lambda_{\min}$ by the normalization of $\rho_C$ and by $\Tr {\bf C}= N$ (the latter requiring $\mu > 1$). Using 
Eq. (\ref{burda}) one can readily compute the dressed spectrum $\rho_E(\lambda)$. In Fig. 2, we show, on top of the empirical and Mar\v{c}enko-Pastur spectrum, the
``bare'' and the dressed power-law spectrum for $\mu=2$. For later convenience, we parameterize the distribution using $\alpha=\lambda_{\min} \in [0,1]$, in which case
$A=(1-\alpha)^2$ and $\lambda_0=2\alpha-1$ (note that $\alpha=1$ corresponds to the Mar\v{c}enko-Pastur case since in this limit $\rho_C(\lambda) = \delta(\lambda-1)$). 
The fit shown in Fig. 2 corresponds to $\alpha=0.35$, and is now very good, suggesting indeed that the correlation of stocks has
a hierarchical structure with a power-law distribution for the size of sectors (on this point, see also \cite{Marsili}). We should point out that a fit using a 
multivariate Student model also works very well for the Pearson estimator of the empirical correlation matrix. However, as noted in \cite{BiroliStudent}, such an
agreement appears to be accidental. If the Student model was indeed appropriate, the spectrum of the most likely correlation matrix (see Eq. (\ref{maxlstudent})) should be
given by Mar\v{c}enko-Pastur, whereas the data does not conform to this prediction \cite{BiroliStudent}. This clearly shows that the Student copula is 
in fact not adequate to model 
multivariate correlations.

\begin{figure}
\begin{center}
\psfig{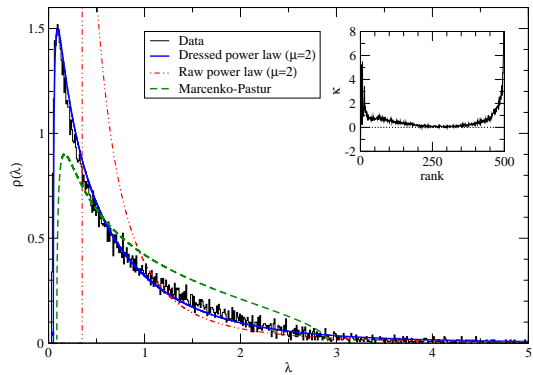} 
\end{center}
\caption{Main figure: empirical average eigenvalues spectrum of the correlation matrix (plain black line), compared to (a) the 
Mar\v{c}enko-Pastur prediction (dashed line) and the dressed power-law spectrum model (thick line). We also show the bare power law 
distribution with $\mu=2$ and the optimal value of $\lambda_{\min}$ (dashed-dotted line). Inset: Kurtosis of the components of the eigenvectors as a function of the 
eigenvalue rank. One clearly sees some structure emerging at both ends of the spectrum, whereas the centre of the band is compatible 
with rotationally invariant eigenvectors.
}
\label{Fig2}
\end{figure}

A complementary piece of information is provided by the statistics of the eigenvectors. Structure-less eigenvectors (i.e. a normalized random vector in $N$ dimensions) have components 
that follow a Gaussian distribution of variance $1/N$. The kurtosis of the components for a given eigenvector gives some indication of its ``non-random'' character (and is trivially
related to the well known inverse participation ratio or Herfindahl index). We show in the inset of Fig. 2 the excess kurtosis as a function of the rank of the eigenvectors (small ranks corresponding
to large eigenvectors). We clearly see that both the largest and the smallest eigenvectors are not random, while the eigenvectors at the middle of the band have a very small excess kurtosis.
As mentioned above, large eigenvalues correspond to economic sectors, while small eigenvalues correspond to long-short portfolios that invest on fluctuations with particularly low volatility, 
for example the difference between two very strongly correlated stocks within the same sector.

\subsection{RMT inspired cleaning recipes}
\label{sect5.2}

As emphasized in Sect. \ref{sect2.3}, it is a bad idea at all to use directly the empirical correlation matrix in a Markowitz optimization program. We have seen that 
the out-of-sample risk is at best underestimated by a factor $(1-q)$, but the situation might be worsened by tail effects and/or by the non-stationarity of the true correlations.
Since we know that measurement noise, induced by the finite size effects, significantly distort the spectrum of the correlation matrix, one should at the very least 
try to account for these noise effects before using the correlation matrix in any optimization program. With the above RMT results in mind, several ``cleaning schemes'' can be 
devised. The simplest one, first suggested and tested in \cite{us2}, is to keep unaltered all the eigenvalues (and the corresponding eigenvectors) that exceed the   
Mar\v{c}enko-Pastur edge $(1 + \sqrt{q})^2$, while replacing all eigenvalues below the edge, deemed as meaningless noise, but a common value $\overline{\lambda}$ such that
the trace of the cleaned matrix remains equal to $N$. We call this procedure eigenvalue clipping, and will consider a generalized version where the $(1-\alpha) N$ largest 
eigenvalues are kept while the $N \alpha$ smallest ones are replaced by a common value $\overline{\lambda}$.

A more sophisticated cleaning is inspired by the power-law distribution model described above. If the true distribution is given by Eq. (\ref{powerlaw}), then we expect 
the $k$th eigenvalue $\lambda_k$ to be around the value:\footnote{Note that actually $\lambda_{\min}$ is given by the very same equation with $k=N$, i.e., it is indeed the smallest 
eigenvalue for large $N$.}
\be\label{lambdak}
\lambda_k \approx \lambda_0 + \left(A \frac{N}{k}\right)^{1/\mu} \longrightarrow_{\mu=2} 2\alpha -1 + (1-\alpha) \sqrt{\frac{N}{k}} 
\ee
The ``power-law'' cleaning procedure is therefore to fix $\mu=2$ and let $\alpha$ vary to generate a list
of synthetic eigenvalues using the above equation Eq. (\ref{lambdak}) for $k > 1$, while leaving the corresponding $k$th eigenvector untouched. Since the market mode $k=1$ is well determined 
and appears not to be accounted for by the power-law tail, we leave it as is.

We will compare these RMT procedures to two classical, so-called shrinkage algorithms that are discussed in the literature (for a review, see \cite{Ledoit}; see also \cite{Guhr,review,Bai} for 
alternative proposals and tests). One is to ``shrink'' the empirical correlation matrix ${\bf E}$ towards the identity matrix:
\be\label{cleaning}
{\bf E} \longrightarrow (1-\alpha) {\bf E}  + \alpha {\bf 1}, \qquad 0 \leq \alpha \leq 1
\ee
An interpretation of this procedure in terms of a minimal diversification of the optimal portfolio is given in \cite{book}. A more elaborate one, due to Ledoit and Wolf, is to replace 
the identity matrix above by a matrix $\overline{\bf C}$ with $1$'s on the diagonal and $\overline{\rho}$ for all off-diagonal elements, where $\overline{\rho}$ is the average of the pairwise correlation
coefficient over all pairs. 

This gives us four cleaning procedures, two shrinkage and two RMT schemes. We now need to devise one or several tests to compare their relative merits. The most natural test that comes to mind is 
to see how one can improve the out-of-sample risk of an optimized portfolio, following the discussion given in Sect. \ref{sect2.3}. However, we need to define a set of predictors we use for 
the vector of expected gains ${\bf g}$. Since many strategies rely in some way or other on the realized returns, we implement the following investment strategy: each day, the empirical correlation 
matrix is constructed using the 1000 previous days, and the expected gains are taken to be proportional to the returns of the current day, i.e. $g_i = r_i^{t}/\sqrt{\sum r_j^{t2}}$. The optimal portfolio with 
a certain gain target is then constructed using Eq. (\ref{Msolution}) with a correlation matrix cleaned according to one of the above four recipes. The out-of-sample risk is 
measured as the realized variance of those portfolios over the next 99 days. More precisely, this reads:
\be
\mathbf{w}^t = \frac{\mathbf{E}^{-1}_\alpha\mathbf{g}^{t}}{\mathbf{g}^{t{{T}}}\mathbf{E}^{-1}_\alpha\mathbf{g}^{t}},
\ee
where $\mathbf{E}_\alpha$ is the cleaned correlation matrix, which depends on a parameter $\alpha$ used in the cleaning algorithm (see for example Eq. (\ref{cleaning}) above). 
A nice property of this portfolio is that if the predictors are normalized by their dispersion on day $t$, the true risk is ${\cal R}^{t2}_\subs{true}=1$. 
The out-of-sample risk is measured as:
\be
{\cal R}^{t2}_\subs{out} = \frac{1}{99} \sum_{t'=t+1}^{t+99} \left[ \sum_i \frac{w_i^t}{\sigma_i^t} r_i^{t'} \right]^2,
\ee
where $\sigma_i^t$ is the volatility of stock $i$ measured over the last 1000 days (the same period used to measure ${\bf E}$). The out-of-sample risk is then averaged over time, and 
plotted in Fig. 3 as a function of $\alpha$ for the four different recipes. In all cases but Ledoit-Wolf, $\alpha=1$ corresponds to the ${\bf E}_\alpha={\bf 1}$ (in the case of the power-law method, $\alpha=1$
corresponds to $\rho_C(\lambda)=\delta(\lambda-1)$). In this case, ${\cal R}^{2}_\subs{out} \approx 25$ which is very bad, since one does not even account for the market mode. 
When $\alpha=0$, ${\bf E}_0$ is the raw empirical matrix, except in the power-law method. 
We show in Fig 3 the in-sample risks as well. From the values found for $\alpha=0$ (no cleaning), one finds that the ratio of out-of-sample to in-sample  risk is $\approx 2.51$, 
significantly worse that the expected result $1/(1-q)=2$. This may be due either to heavy tail effects or to non stationary effects (see next subsection). 
The result of Fig. 3 is that the best cleaning scheme (as measured from this particular test) is eigenvalue clipping, followed by the power-law method. 
Shrinkage appears to be less efficient than RMT-based cleaning; this conclusion is robust against changing the quality factor $Q$. 
However, other tests can be devised, that lead to slightly different conclusions. One simple variant of the above test is to 
take for the predictor ${\bf g}$ a random vector in $N$ dimensions, uncorrelated with the realized returns. Another idea is to use to correlation matrix to define residues, i.e. how well the returns of a
given stock are explained by the returns of all other stocks on the same day, excluding itself. The ratio of the out-of-sample to in-sample residual variance is another measure of the quality of the cleaning. 
These two alternative tests are in fact found to give very similar results. The best cleaning recipe now turns out to be the power-law method, while the eigenvalue clipping is the worst one. Intuitively,
the difference with the previous test comes from the fact that random predictor ${\bf g}$ is (generically) orthogonal to the top eigenvectors of 
${\bf E}$, whereas a predictor based on the returns 
themselves has significant overlap with these top eigenvectors. Therefore, the correct description of the corresponding eigenvalues is more important in the latter case, 
whereas the correct treatment of strongly correlated pairs (corresponding to small eigenvectors) is important to keep the residual 
variance small.

\begin{figure}
\begin{center}
\psfig{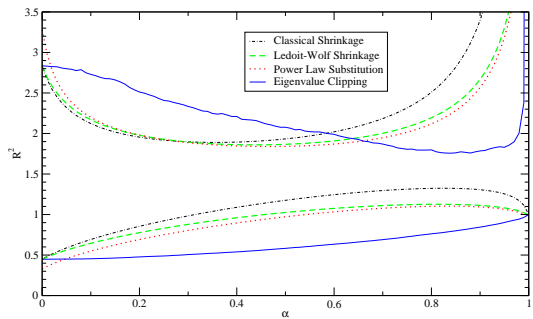} 
\end{center}
\caption{Comparison between different correlation matrix cleaning schemes for Markowitz optimization. Top curves: out-of-sample squared risk 
${\cal R}^{2}_\subs{out}$ as a function of the cleaning parameter $\alpha$ (see Eq. \ref{cleaning}). $\alpha=0$ corresponds to the raw empirical
correlation matrix, and $\alpha=1$ to the identity matrix. The best cleaning correspond to the smallest out-of-sample risk. 
The `true' risk for this problem is ${\cal R}_\subs{true}=1$. Bottom curves: in-sample risk of the optimal portfolio as a function of $\alpha$.
}
\label{Fig3}
\end{figure}

In summary, we have found that RMT-based cleaning recipes are competitive and outperform, albeit only slightly, more traditional shrinkage algorithms when applied to portfolio optimization or 
residue determination. However, depending on the objective and/or on the structure of the predictors, the naive eigenvalue clipping method proposed in \cite{us2} might not be appropriate. 
In view of both the quality of the fit of the eigenvalue distribution (Fig. 2) and the robustness of the results to a change of the testing method, our
analysis appears overall to favor the power-law cleaning method. However, one should keep in mind that the simple minded shrinking with $\alpha=1/2$ is quite robust and in fact difficult to beat, 
at least by the above RMT methods that do not attempt to mop up the eigenvectors.

\subsection{Dynamics of the top eigenvector -- an empirical study}
\label{sect5.3}

Finally, we investigate possible non stationary effects in financial markets by studying the dynamics of the top eigenvalue and eigenvector. In 
order to even measure these quantities, one needs a certain averaging time scale, noted $1/\epsilon$ in Sect. \ref{sect4.4} above. If the true top
eigenvector (or eigenvalue) did not evolve in time, the variograms defined in Eqs. (\ref{varval},\ref{varvec}) should converge to their asymptotic limits 
after a time $\tau \sim \epsilon^{-1}$. If the structure of the correlations does really change over long times, there should be a second relaxation mode for these quantities,
leading to an increased asymptotic value for the variograms with, possibly, a slower relaxation mode contributing to a long time tail in the variogram.
Empirical evidence for such a long term evolution of the market mode was presented in \cite{Krakow1}. Here, we 
repeat the analysis on the above data set, with now a fixed pool of stocks containing 50 stocks for the whole period. The time scale $1/\epsilon$ is chosen to be $25$ days. In Fig. 4 we show the variograms where one clearly see the existence of genuine fluctuations of the 
market mode on a time scale $\sim 100$ days, superimposed to the initial noise dominated regime that should be described by Eqs. (\ref{varval},\ref{varvec}). The asymptotic value of 
these variograms  is furthermore much larger than predicted by these equations. In particular, the variogram of the largest eigenvector should converge to $\approx 0.08$.
One should thus expect that the `true' correlation matrix ${\bf C}$ is indeed time dependent with a relatively slow evolution on average
(although correlation `crashes' have been reported). This genuine non-stationarity of financial markets is, to some extent, expected. \footnote{Some instruments actually price the `implied correlation', i.e. the average correlation between stocks expected by the market over some 
future time period!} It makes quantitative modelling difficult and sometimes even dangerous; even if a perfect cleaning scheme was available, the out-of-sample risk would always be underestimated. New ideas to understand, model and predict the correlation dynamics are clearly needed. 

\begin{figure}
\begin{center}
\psfig{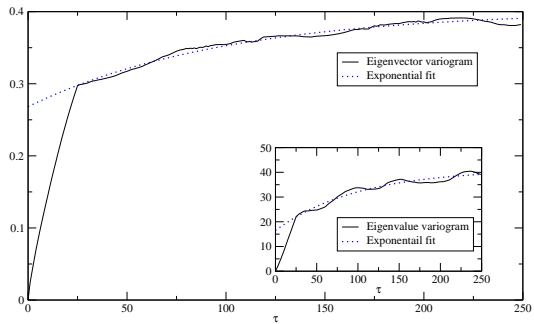} 
\end{center}
\caption{Variogram of the top eigenvector, defined by Eq. (\ref{varvec}) (main plot) and of the corresponding eigenvalue, Eq. (\ref{varvec}) (inset). 
The long term evolution is fitted by an exponential relaxation with a characteristic time around 100 days. Since the time periods are non overlapping, 
the value for $\tau=25$ days should correspond to the asymptotic values in Eqs. (\ref{varval},\ref{varvec}), but the latter are much smaller than the 
empirical values found here. These features show that the structure of the correlation matrix is itself time dependent.
}
\label{Fig4}
\end{figure}

\section{Some open problems}
\label{sect6.0}
\subsection{Cleaning the eigenvectors?}

As we reviewed in the previous section, RMT has already significantly contributed to improving the reconstruction of the correlation 
matrix from empirical data. However, progress is limited by the fact that most RMT results concern eigenvalues but say little about 
eigenvectors. It is not obvious to formulate natural priors for the structure of these eigenvectors -- from symmetry alone, one can 
only argue that the top eigenvalue of the correlation matrix of an ensemble of stocks should be uniform, but even this is not 
obvious and there is a clear market capitalization dependence of the weights of the empirical top eigenvector. In order to make
headway, one should postulate some a priori structure, for example factor models, or ultrametric tree models \cite{Lillo3}. Whereas our knowledge of 
the influence of noise on the eigenvalue spectrum is quite satisfactory, the way `true' eigenvectors are dressed by measurement noise is to a large extent 
unexplored (the case of a well separated top eigenvalue was treated in Sect. \ref{sect4.4} above). Statistical techniques to ``clean''  
eigenvectors with a non trivial structure are needed (for a very recent attempt, see \cite{NewNYU}). As a matter of fact, results concerning the structure of 
eigenvectors are difficult as soon as one walks away from the assumption of statistical invariance under orthogonal transformations. For example, 
the structure of the eigenvectors of L\'evy matrices is extremely rich and numerically display interesting localization transitions \cite{CB}. However,
analytical results are scarce. 

\subsection{Time and frequency dependent correlation matrices}

In order to guess correctly the structure of correlations in financial markets, it seems important to understand how these correlations appear from
the high frequency end. It is clear that prices change because of trades and order flow. Correlations in price changes reflect correlations in order flow.
Detailed empirical studies of these order flow correlations at the tick by tick level are not yet available, but important progress should be witnessed soon.
On a more phenomenological level, one can gain intuition by postulating that the way stock $i$ moves between $t$ and $t+{\rm d}t$, measured by the
return $r_i^t$, depends on the past returns of all other stocks $j$. If one posits a causal, linear regression model for the lagged 
cross-influences, one may write \cite{Krakow1}:
\be 
r_{i}^t = \xi_i^t + \sum_j \int_{-\infty}^{+\infty} dt' K_{ij}(t-t') r_{j}^{t'}; \qquad {\mbox{with}} \quad K_{ij}(\tau < 0) \equiv 0
\ee
where $\xi_i$ represents the idiosyncratic evolution of stock $i$, due to the high frequency component of order flow. For ${\rm d}t \to 0$, on may 
assume for simplicity that these components are uncorrelated in time, i.e.:
\be
\langle  \xi_i^t \xi_j^{t'} \rangle = C_{ij}^0 \delta_{ij} \delta(t-t'),
\ee
where $C_{ij}^0$ is the high frequency ``bare'' correlation matrix, that come from simultaneous trading of different stocks. 
The matrix $K_{ij}$ describe how the past returns of stock $j$ drive those of stock $i$. The $K_{ij}$ can be thought of as ``springs'' that hold the price
of different stocks together. 

Strongly correlated pairs of stocks are described by a strong cross-influence term $K_{ij}$. Presumably some stocks are `leaders' while other, smaller
cap stocks, are laggers; this means that in general $K_{ij}(\tau) \neq K_{ji}(\tau)$. Denoting the Fourier transform of the lag dependent correlation $C_{ij}(\tau)$ (defined by Eq. (\ref{Ctau})) as $\widehat C_{ij}(\omega)$, one finds:
\be
\widehat C_{ij}(\omega) = \sum_{kk'} (1 - K(\omega))^{-1}_{ik} C_{kk'}^0 (1 - K(-\omega))^{-1}_{jk'}.
\ee
This model suggests that, arguably, the kernels $K_{ij}(\tau)$ captures more directly the microscopic mechanisms that construct the low frequency correlation matrix and is a 
fundamental object that one should aim at modelling, for example to infer meaningful influence networks. 
The correlation matrix reflects the way people trade, and possibly the correlation between their 
trading strategies on different stocks. Models that explicitly take into account this feedback mechanism between ``bare'' correlations and the impact of trading 
only start to be investigated \cite{Wyart,Marsili2,Rama}, and appear to be particularly fruitful to understand the issue of non-stationarity, i.e. how the correlation matrix 
itself may evolve in time, for example in crisis periods. More work in that direction would certainly be interesting, because gradual or sudden changes of the correlation matrix is, as noted above, an important concern in risk management.

\subsection{Non-linear correlations and copulas}

We have mentioned above that a full characterization of the correlation structure amounts to specifying a ``copula''. In particular, non-linear correlations, 
such as $\langle r_i^2 r_j^2 \rangle_c$, or tail correlations, can be very important for some applications (for example, the risk of an option portfolio). 
We have seen that the Student copula, which is intuitive and popular, in fact fails to describe stock data. The construction of an adequate model for the linear and 
non-linear correlations of stocks is still very much an open problem.

\subsection{Random SVD and Canonical Component Analysis}

Finally, let us mention two problems that may be worth considering, concerning the problem of random singular value decomposition. We have seen that in the case
where there is no correlation whatsoever between $N$ input and $M$ output variables, the spectrum of empirical singular values with $T$ observations is given by
Eq. (\ref{resultSVD}), which is the analogue of the Mar\v{c}enko-Pastur result. In practical cases, however, there might be some true correlations between input and
output variables, described by a non-trivial `true' spectrum of singular values. The modification of this spectrum due to measurement noise (i.e. the analogue 
of Eq. (\ref{burda_basic})) is, to our knowledge, not known. A detailed analysis of the influence of heavy tails in the distribution of both input and
output variables would also be quite satisfying.

\subsection*{Acknowledgements} We want to warmly thank all our collaborators on these topics, in particular Giulio Biroli, Laurent Laloux and
Augusta Miceli. We also acknowledge many useful conversations, over the years, with G\'erard Ben Arous, Zadislaw Burda, Stefano Ciliberti, 
Iain Johnstone, Imre Kondor, Fabrizio Lillo, Satya Majumdar, Matteo Marsili and Sandrine P\'ech\'e.

\end{document}